\documentclass[onecolumn,secnumarabic,amsmath,amssymb,balancelastpage,nofootinbib]{article}

\usepackage[e]{esvect}
\usepackage{color}         
\usepackage{graphics}      
\usepackage{graphicx}      
\usepackage{epsf}          
\usepackage{bm}            

\usepackage{cite}
\usepackage{amssymb}
\usepackage{amsmath, esint}
\usepackage{mathrsfs}
\usepackage{framed}
\usepackage{bigints} 
\usepackage{enumitem}
\usepackage{pifont}
\usepackage{setspace} 

\usepackage[none]{hyphenat} 

\usepackage[colorlinks=true]{hyperref}  

\definecolor{darkred}{rgb}{0.6,0,0}
\definecolor{darkgreen}{rgb}{0,0.5,0}
\definecolor{darkblue}{rgb}{0,0,0.6}
\hypersetup{ colorlinks,
linkcolor=darkblue,
filecolor=darkgreen,
urlcolor=darkgreen,
citecolor=darkred }

\begin{document}

\sloppy 

\bibliographystyle{nar}

\newlength{\bibitemsep}\setlength{\bibitemsep}{.2\baselineskip plus .05\baselineskip minus .05\baselineskip}
\newlength{\bibparskip}\setlength{\bibparskip}{0pt}
\let\oldthebibliography\thebibliography
\renewcommand\thebibliography[1]{%
  \oldthebibliography{#1}%
  \setlength{\parskip}{\bibitemsep}%
  \setlength{\itemsep}{\bibparskip}%
}


\title{\vspace*{-44 pt}\Huge{Electron Charge Density:}\\\huge{A Clue from Quantum Chemistry\\for Quantum Foundations}}
\author{Charles T. Sebens\\California Institute of Technology}
\date{\vspace*{-6 pt}arXiv v.2\ \ \ \ June 24, 2021\vspace*{10 pt}\\Forthcoming in \textit{Foundations of Physics}}

\maketitle
\vspace*{-20 pt}
\begin{abstract}	
Within quantum chemistry, the electron clouds that surround nuclei in atoms and molecules are sometimes treated as clouds of probability and sometimes as clouds of charge.  These two roles, tracing back to Schr\"{o}dinger and Born, are in tension with one another but are not incompatible.  Schr\"{o}dinger's idea that the nucleus of an atom is surrounded by a spread-out electron charge density is supported by a variety of evidence from quantum chemistry, including two methods that are used to determine atomic and molecular structure: the Hartree-Fock method and density functional theory.  Taking this evidence as a clue to the foundations of quantum physics, Schr\"{o}dinger's electron charge density can be incorporated into many different interpretations of quantum mechanics (and extensions of such interpretations to quantum field theory).
\end{abstract}

\tableofcontents
\newpage

\section{Introduction}\label{introsec}

Despite the massive progress that has been made in physics, the composition of the atom remains unsettled.  J. J. Thomson \cite{thomson1904} famously advocated a ``plum pudding'' model where electrons are seen as tiny negative charges inside a sphere of uniformly distributed positive charge (like the raisins---once called ``plums''---suspended in a plum pudding).  This model was quickly refuted by the Rutherford-Geiger-Marsden metal foil experiments, but there is disagreement as to what should take its place.  The positively charged pudding has been compressed to a dense nucleus at the center of the atom, but the fate of the negatively charged raisins is disputed.  Often, the atom is described as mostly empty space with a few loose raisins flitting about.  However, there is another option: we can take the familiar electron cloud of an atom to be a cloud of negative charge surrounding the positively charged nucleus.  This picture is not as conducive to culinary metaphors, but we might flip the plum pudding model and think of the electrons as the pudding and the nucleus as a cluster of positively charged (and neutral) raisins stuck together at the center of a poorly crafted plum pudding.

The view of electrons as pudding spread-out across atoms and molecules often crops up in quantum chemistry, though it is not always clearly distinguished from the alternative view of electrons as raisins.  A welcome exception is Bader and Matta \cite{bader2013}:
\begin{quote}
``The electronic charge, unlike that of the more massive nuclei, is spread throughout space, and matter is made up of point-like nuclei embedded in the relatively diffuse spatial distribution of electronic charge. The distribution of electronic charge is described by the electron density that determines the amount of negative charge per unit volume.'' \cite[pg.\ 255]{bader2013}
\end{quote}
The idea here is that the amplitude-squared of the quantum wave function gives the density of electron charge (exactly how will be explained in section \ref{PCsection}).  That idea can be traced back to Schr\"{o}dinger in his papers introducing the Schr\"{o}dinger equation.  This role for amplitude-squared in determining charge density is sometimes contrasted with Born's role for amplitude-squared in determining probabilities for measurement outcomes.  In section \ref{PCsection}, we will see that the amplitude-squared could potentially play both roles: either because the density of charge contracts (or at least appears to) during quantum measurements or because Schr\"{o}dinger's density of charge is not an instantaneous charge density (electrons being rapidly moving raisins that look like pudding when blurred over a short period of time).

Physicists and philosophers have put forward a number of interpretations of quantum mechanics that seek to be clear about the ontology of the theory (what exists according to the theory), among which we will focus here on Ghirardi-Rimini-Weber (GRW) theory, the many-worlds interpretation, Bohmian mechanics, and the recent proposals by Hall \emph{et al.}\ \cite{HDW}, Sebens \cite{sebens2015}, and Gao \cite{gao2014, gao2017, gao2018, gao2020}.  Such attempts to clarify the ontology of quantum mechanics do not always include a charge density like the one Schr\"{o}dinger proposed, though we will see in section \ref{QFsection} that it can often be added.  Recognizing that such a charge density can be included, one might wonder whether it should be.  What evidence do we have that, at least in atoms and molecules, Schr\"{o}dinger's electron charge density accurately describes the distribution of electron charge?

In section \ref{QCsection}, I will present (suggestive, but not definitive) evidence from two methods that are used to approximate the shapes of electron clouds in atoms and molecules, the arrangement of nuclei in molecules, and the ground state energies of atoms and molecules: the Hartree-Fock method and density functional theory.  In both methods, the ground state potential energies of atoms and molecules can be divided into classical contributions (that can be calculated from Schr\"{o}dinger's electron charge density) and quantum corrections.  The Hartree-Fock method and density functional theory are both widely applied textbook techniques in quantum chemistry, but I will present them in detail as they may not be familiar to all working in the foundations of quantum mechanics.  There is a tendency to skip over such approximation techniques when attempting to understand the theory's foundations.  However, I think there is much to be learned about quantum mechanics by studying the approximations that are used to model atoms and molecules.

I see the evidence for a spread-out electron charge density from quantum chemistry as evidence ``from above'' non-relativistic quantum mechanics, as it is evidence from the way the theory is approximated and applied.  Although it is not a focus of this article, I also believe there is strong evidence ``from below,'' from the deeper physics that non-relativistic quantum mechanics approximates: relativistic quantum field theory.  As will be discussed briefly in section \ref{QFT}, quantum field theory can be viewed as a theory of fields in quantum superpositions of classical states where the classical field states have spread-out charge distributions.

\section{Probability Density and Charge Density}\label{PCsection}

Within quantum chemistry, the squared amplitude of a single-electron wave function plays two distinct roles.  First, it plays the standard role---dating back to Born's 1926 article \cite{born1926}---of giving the probability density for detecting the particle in a given region of space.  Second, when multiplied by the charge of the electron, $-e$, the squared amplitude gives the density of charge.  This second role was proposed by Schr\"{o}dinger \cite{schrodinger1926pt3, schrodinger1926pt4, schrodinger1926rev, schrodinger1928report, schrodingerletter} before Born proposed the first role.  When this historical episode is presented in textbooks on quantum mechanics or quantum chemistry, we are usually told that it was Born (not Schr\"{o}dinger) who correctly understood the role of amplitude-squared.\footnote{This is also how Born \cite{born1955} retells the story.  Other authors explicitly challenge the common narrative and defend the potential value of Schr\"{o}dinger's original charge density role (e.g., Einstein \cite[pg.\ 168--169]{einstein1934}; Jaynes \cite{jaynes1973}; Bader \cite{bader1990, bader2003, bader2010}; Bell \cite[pg.\ 39--40]{bell1990}; Bacciagaluppi and Valentini \cite[ch.\ 4]{bacciagaluppi2009}; Allori \emph{et al.}\ \cite{allori2011}; Norsen \cite[ch.\ 5]{norsen2017}; Gao \cite{gao2017, gao2018}).}  Here is a representative quotation\footnote{See also \cite[sec.\ 3.6]{gillespie2001}; \cite[ch.\ 17]{longair2013}; \cite[pg.\ 10, 147, 460]{levineQC}.} taken from a discussion of the one-dimensional infinite square well in McQuarrie's quantum chemistry textbook:
\begin{quote}
``...Schr\"{o}dinger considered [$-e$]$\psi^*(x)\psi(x)$ to be the charge density and [$-e$]$\psi^*(x)\psi(x)dx$ to be the amount of charge between $x$ and $x+dx$.  Thus, he presumably pictured the electron to be spread all over the region [between the barriers of the well].  A few years later, however, Max Born ... found that this interpretation led to logical difficulties and replaced Schr\"{o}dinger's interpretation with $\psi^*(x)\psi(x)dx$ as the \emph{probability that the particle is located between $x$ and $x+dx$}.'' \cite[pg.\ 104]{mcquarrieQC}
\end{quote}
Still, despite the problems facing Schr\"{o}dinger's proposal (some of which will be discussed later, in section \ref{QFsection}), his idea that $-e$ times the amplitude-squared of an electron's wave function gives the electron's charge density has not been fully rejected.  In fact, when describing atoms and molecules, the authors of quantum chemistry textbooks often move from probability density to charge density without comment, as if it is obvious that they should be proportional to one another.  For example, Szabo and Ostlund write:\footnote{Similar examples appear in \cite[pg.\ 151]{szaboQC}; \cite[pg.\ 6]{bader1990}; \cite[pg.\ 771]{shusterman1997}; \cite[pg.\ 1141]{matta2002}; \cite[pg.\ 223]{atkins2011}; \cite[pg.\ 403, 460]{levineQC}.}
\begin{quote}
``If we have an electron described by the spatial wave function $\psi_a(\bm{r})$, then the probability of finding that electron in a volume element $d\bm{r}$ at a point $\bm{r}$ is $|\psi_a(\bm{r})|^2 d\bm{r}$.  The probability distribution function (charge density) is $|\psi_a(\bm{r})|^2$.'' \cite[pg.\ 138]{szaboQC}
\end{quote}
Speaking carefully, the charge density would really be $-e$ times the amplitude-squared.  However, some authors omit this constant and leave it implicit, calling the amplitude-squared itself a ``charge density'' (e.g., Bader \cite{bader1990}).

\begin{figure}[htb]
\center{\includegraphics[width=13 cm]{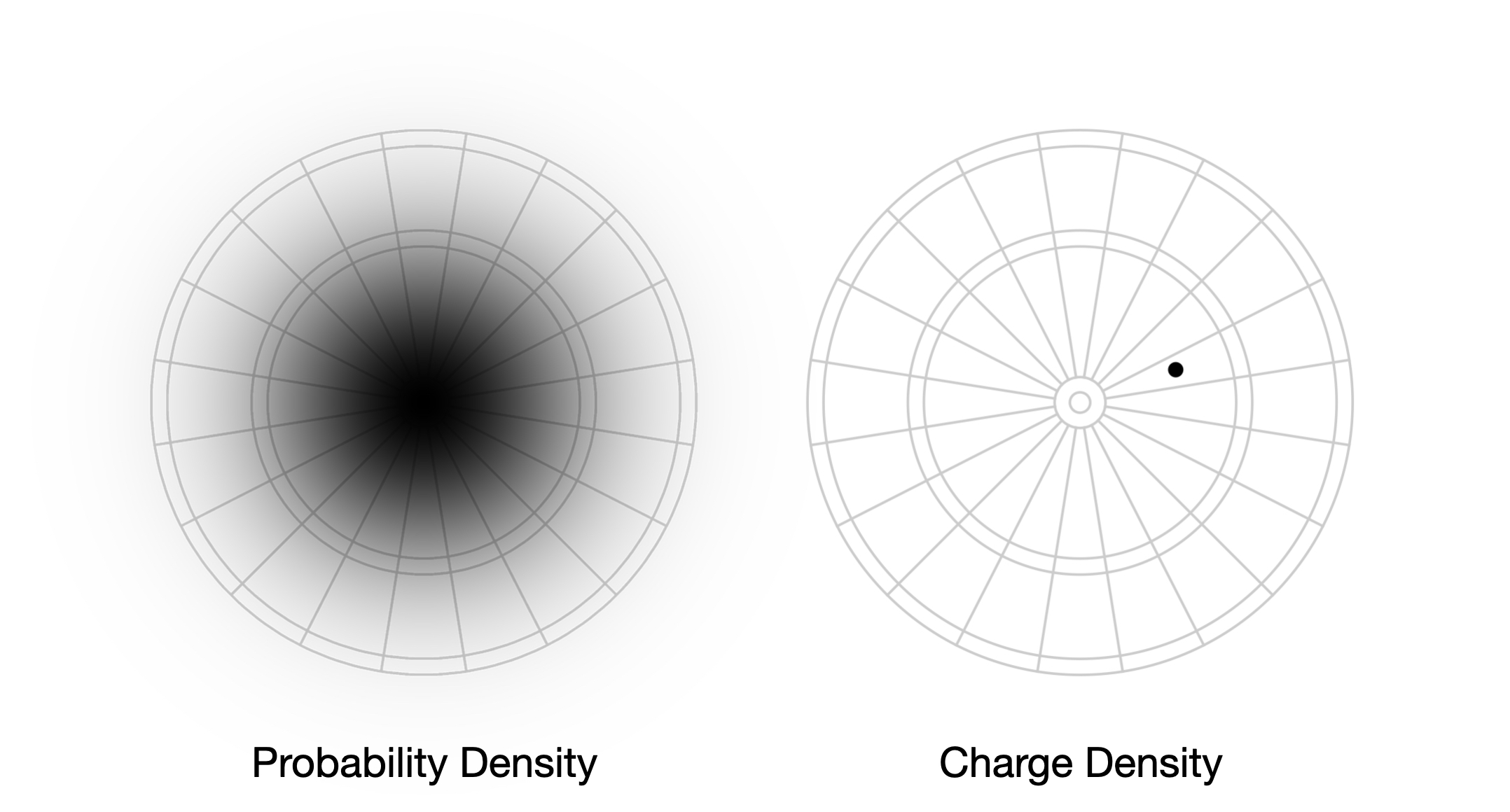}}
\caption{For a charged dart thrown at a dartboard, the probability distribution is shown as a black cloud centered at the bullseye.  Even though the probability distribution is spread out, the charge distribution is not.  The dart's entire charge is concentrated at the location where it actually hits the board.}
\label{darts}
\end{figure}

The two quotes above differ on whether they say that the probability density gives probabilities for the electron being located in certain regions at this moment or only claim that it gives probabilities for the electron being found in certain regions upon measurement.  To see where it leads us, let us adopt the former view and take the probability density to give probabilities for where the electron is now.  Understanding probability density in this way, the amplitude-squared cannot play both of the roles described above.  We must choose between Born and Schr\"{o}dinger.  As an analogy, consider throwing a small charged dart at a large non-conducting dartboard (figure \ref{darts}).  After tossing, the probability density for the dart being at a given location on the board might be a two-dimensional Gaussian (of some specified width) centered at the  bullseye.  But, that is not an accurate representation of the dart's charge density.  The charge is confined to the dart's actual location, wherever that may be.  Similarly, if the electron is a point charge with a precise location, then its charge density is a delta function centered at the electron's location.  It is not given by $-e$ times the amplitude-squared of the electron's wave function.

In the next section, we will see that there are good reasons to take Schr\"{o}dinger's idea of an electron charge density seriously in quantum chemistry.  If we want to hold on to it, we need to somehow resolve the above tension.  There are a number of strategies that could be pursued.

One idea is that the kind of charge density that is proportional to amplitude-squared does not describe the actual distribution of charge but instead describes probabilistically where the electron might be (viewing the electron as a point charge).  Being precise, we should call this the \emph{expected charge density} to distinguish it from the actual charge density (which is a delta function centered at the point electron's location).  The expected charge density at a point in space is not the charge density we actually expect to see, but instead a probabilistically weighted summary of our expectations about where the electron might be (given its wave function).  Similarly, in the dart analogy, we do not expect to see a spread-out charge density even though the expected charge density is spread out.  This way of retaining a kind of charge density that is proportional to amplitude-squared amounts to declaring complete victory for Born and is insufficient for our purposes.  As we will see, the evidence from quantum chemistry does not point to the electron's charge merely being possibly located in different places.  It points instead to there being an actual spread-out distribution of electron charge.

A second idea is that the point electron moves rapidly throughout its wave function so that if we average the charge density over a short time interval it will be approximately proportional to the amplitude-squared---even though at any particular moment the electron's charge density is a delta function centered at the electron's location.  This modification of Schr\"{o}dinger's original picture of a spread-out electron seems to be what Levine has in mind when he writes in his quantum chemistry textbook:
\begin{quote}
``$\rho$ is the electron probability density.  The corresponding charge density averaged over time is equal to $-e\rho(x,y,z)$, where $-e$ is the charge of the electron.'' \cite[pg.\ 403]{levineQC}
\end{quote}
Why do we need to average over time?  Levine seems to be imagining that at any moment the point electron is at a particular location and thus that it is only by blurring over a period of time that you could have a charge density proportional to the probability density at a particular moment.\footnote{Szabo and Ostlund \cite[sec.\ 3.1.1]{szaboQC} sound like they may be thinking along the same lines when they contrast the Coulomb potential sourced by an electron's ``instantaneous position'' with the Coulomb potential arrived at by taking a psi-squared-weighted average over the electron's possible positions.  Nelson \cite{nelson1990} compares the above understanding of amplitude-squared as describing time-averaged particle motion to a number of alternatives.}  Levine is more explicit about this vision when explaining the Hellmann-Feynman electrostatic theorem (which will be discussed briefly in section \ref{FEsection}):
\begin{quote}
``The rapid motion of electrons causes the sluggish nuclei to `see' the electrons as a charge cloud rather than as discrete particles.'' \cite[pg.\ 430]{levineQC}
\end{quote}
Pauling presents a similar picture when discussing the ground state of the hydrogen atom, writing:
\begin{quote}
``Over a period of time long enough to permit many cycles of motion of the electron the atom can be described as consisting of the nucleus surrounded by a spherically symmetrical ball of negative electricity (the electron blurred by a time-exposure of its rapid motion), as indicated in [a figure depicting the electron cloud].'' \cite[pg.\ 15]{pauling1960}
\end{quote}
Returning to the analogy with the dart and stretching it a bit, the idea here is that the dart zips all over the board so that over any short time interval it explores every section of the board and spends more time near the bullseye than far from it.

To fill out this picture, we would need a dynamical theory of how the point electron moves through its wave function.  Bohmian mechanics is the most thoroughly developed formulation of quantum mechanics that gives such a dynamics.  But, Bohmian mechanics does not vindicate the picture that Levine and Pauling paint (see section \ref{BMsection}).  Still, one may seek an alternative dynamics for which $-e$ times the amplitude-squared of the electron wave function gives the time-averaged charge density (an approach recently advocated by Gao \cite{gao2014, gao2017, gao2018, gao2020}).

A third idea for reconciling Born and Schr\"{o}dinger's roles is to retreat to the weaker version of Born's probabilistic role for amplitude-squared, taking the probability density to give probabilities for future measurements of the electron's location.  This then allows one to consistently hold on to both Born and Schr\"{o}dinger's roles for amplitude-squared (without modifying Schr\"{o}dinger's role in the way Levine and Pauling do above), though we must give up the point electron.  On this view, before measurement the electron's charge is distributed throughout its wave function and after measurement it is concentrated wherever the electron is found.  Returning to the dart analogy, imagine that instead of throwing a dart you throw a sticky blob of charged goo that splats with its mass and charge spread over a Gaussian centered at the bullseye.  Then, when you approach the board and perform a measurement to locate the goo, the interaction causes it to concentrate in a small region.

As will be discussed in section \ref{QFsection}, the idea of a charge density that becomes more concentrated upon measurement fits well with formulations of quantum mechanics that modify the dynamics to include collapse of the wave function (such as GRW theory).  It can also be incorporated into theories that include only apparent collapse, like the many-worlds interpretation.  Although this picture of a spread-out charge density does not match ordinary Bohmian mechanics, it fits well with certain Bohmian approaches to quantum field theory that take fields to be more fundamental than particles.  In fact, we'll see that this picture of spread-out charge density can be combined with a variety of approaches to quantum field theory.

Thus far, we have kept things simple by focusing on single-electron wave functions.  For multiple electrons, one cannot simply say that charge density is proportional to the wave function's amplitude-squared because the wave function is no longer a function of just one spatial location.  Instead, the charge density that was proposed by Schr\"{o}dinger \cite{schrodinger1926pt4, schrodingerletter}---and is used now\footnote{In quantum chemistry, see (for example) \cite[sec.\ 3.8.3]{szaboQC}; \cite[sec.\ 1.3]{bader1990}; \cite{bader2010}; \cite{matta2002}; \cite[pg.\ 403]{levineQC}.}---involves summing a contribution from each electron where the coordinates of the others are integrated out.  If we temporarily set aside electron spin, this charge density is given by the expression
\begin{align}
\rho^q(\vec{x})&=(-e) \int{|\psi(\vec{x},\vec{x}_2, \vec{x}_3, ...)|^2 \ d^3 x_2 d^3 x_3 ...}
\nonumber
\\
&\quad + (-e) \int{|\psi(\vec{x}_1,\vec{x}, \vec{x}_3, ...)|^2 \ d^3 x_1 d^3 x_3 ...}
\nonumber
\\
&\quad + ...
\ ,
\label{sumchargedensity}
\end{align}
where the $q$ superscript on $\rho^q(\vec{x})$ indicates that this is a density of charge.  Because of the antisymmetry of multi-electron wave functions, each term in \eqref{sumchargedensity} makes the same contribution to the total.  Thus, we can write this charge density more concisely as
\begin{equation}
\rho^q(\vec{x})=(-N e) \int{|\psi(\vec{x},\vec{x}_2, \vec{x}_3, ...)|^2 \ d^3 x_2 d^3 x_3 ...}
\ ,
\label{totalchargedensity}
\end{equation}
where $N$ is the total number of electrons.  As the wave function evolves, the total charge---found by integrating \eqref{totalchargedensity} over all space---will always remain $-Ne$.  In the equations above and below, we leave out the time dependence and focus on representing wave functions and charge densities at a single moment.

Adopting Born's role for amplitude-squared, we can divide the expression for charge density in \eqref{sumchargedensity} by $-e$ to yield a function that assigns to each point in space the sum of the probabilities for finding each electron there,
\begin{align}
&\int{|\psi(\vec{x},\vec{x}_2, \vec{x}_3, ...)|^2 \ d^3 x_2 d^3 x_3 ...} + \int{|\psi(\vec{x}_1,\vec{x}, \vec{x}_3, ...)|^2 \ d^3 x_1 d^3 x_3 ...}+ ...
\nonumber
\\
&\quad\quad = N \int{|\psi(\vec{x},\vec{x}_2, \vec{x}_3, ...)|^2 \ d^3 x_2 d^3 x_3 ...}
\ .
\label{numdensity}
\end{align}
This is sometimes called the ``electron number density,'' or simply the ``electron density.''  If electrons are spread out in the way that Schr\"{o}dinger suggested, then \eqref{numdensity} can be interpreted as an actual number density where, for example, each octant of a helium atom would only contain a quarter of an electron.  However, if we are focusing on Born's probabilistic role for amplitude-squared, then the terms ``number density'' and ``electron density'' should be interpreted as shorthand for ``expected number density.''  Integrating \eqref{numdensity} over a region gives an expectation value for the number of electrons in that volume.  In the dart analogy, we might imagine throwing five darts where we know the joint probability distribution for the darts hitting at each location on the board.  Summing probability densities for each dart, as in \eqref{numdensity}, would yield an expected number density spread all over the board---perhaps a two-dimensional Gaussian.  With just five darts, there is no way to have the actual number density resemble a two-dimensional Gaussian.  With a vast number of darts, the actual number density could resemble a two-dimensional Gaussian and thus could resemble the expected number density.

The two potentially viable ideas discussed above for reconciling Born and Schr\"{o}dinger's roles for amplitude-squared can be applied straightforwardly to multiple electron wave functions.  Just as one electron could zip around space so that the time-averaged charge density is $-e|\psi|^2$, many electrons could swarm around so that the time-averaged charge density matches \eqref{totalchargedensity}.  Alternatively, we could say that the electric charge of the electrons is spread out in accord with \eqref{totalchargedensity} and that (upon integration) \eqref{numdensity} yields the expected number of electrons that would be found in a given region were we to perform the appropriate measurement.

Before moving on to the next section, let us generalize the above expressions for charge density and expected number density to account for the fact that electrons have spin.  Starting with a single electron, we move from a single component wave function $\psi(\vec{x})$ to a two component wave function,
\begin{equation}
\chi(\vec{x})=\left(
\begin{matrix}
\psi(\vec{x},1) \\
\psi(\vec{x},2)
\end{matrix}
\right)
\ .
\end{equation}
The charge density for this single-electron wave function is
\begin{equation}
\rho^q(\vec{x})=(- e) \chi^{\dagger}(\vec{x})\chi(\vec{x}) =\sum_{s=1}^{2}(- e) |\psi(\vec{x},s)|^2
\ ,
\label{chargedensitypauli}
\end{equation}
where $s$ is a spin index distinguishing the two components of $\chi(\vec{x})$.  To keep the language easy to follow, I will call $\chi^{\dagger}\chi$ the ``amplitude-squared'' of the wave function even though it is really the sum of the squared amplitudes of each of the two components of $\chi$.  For multiple particles, the charge density is
\begin{equation}
\rho^q(\vec{x})=(-N e) \sum_{s_1=1}^{2}\sum_{s_2=1}^{2}\sum_{s_3=1}^{2}...\int{|\psi(\vec{x},s_1,\vec{x}_2,s_2,\vec{x}_3,s_3,...)|^2\ d^3 x_2 d^3 x_3 ...}
\ .
\label{totalchargedensitypauli}
\end{equation}
As in \eqref{numdensity}, the expected number density differs by a factor of $-e$,
\begin{equation}
N \sum_{s_1=1}^{2}\sum_{s_2=1}^{2}\sum_{s_3=1}^{2}...\int{|\psi(\vec{x},s_1,\vec{x}_2,s_2,\vec{x}_3,s_3,...)|^2\ d^3 x_2 d^3 x_3 ...}
\ .
\label{numdensitypauli}
\end{equation}

To review, for multiple electrons we can distinguish Schr\"{o}dinger's role for amplitude-squared from Born's as follows:
\begin{description}
\item [Schr\"{o}dinger's Role]  The actual electron charge density for a multi-electron wave function is given by \eqref{totalchargedensitypauli} (or by \eqref{totalchargedensity} if we set aside spin).
\item [Born's Role]  The expected electron number density (probabilistically describing the results of possible measurements) for a multi-electron wave function is given by \eqref{numdensitypauli} (or by \eqref{numdensity} if we set aside spin).
\end{description}
The title ``Born's role'' sounds very much like ``Born's rule,'' but a different name is needed because this role is only a special case of the more general Born rule used in contemporary quantum mechanics to generate probabilities for various different kinds of measurements.  To state Schr\"{o}dinger's role precisely, we would want to specify whether the electron charge density in \eqref{totalchargedensitypauli} is an instantaneous charge density or a time-averaged charge density.  For now, let us leave both options on the table.  Schr\"{o}dinger \cite{schrodinger1926pt4} also proposed a current density to describe the flow of charge, but we will not be examining that density here (see \cite{bader2010}).

When Schr\"{o}dinger and Born's roles are contrasted, we are usually advised to abandon Schr\"{o}dinger's role and use only Born's.  However, we will see in the next section that Schr\"{o}dinger's role for amplitude-squared should not be so quickly dismissed.  Schr\"{o}dinger's idea that electron charge is distributed in accordance with amplitude-squared is supported by the techniques currently used to calculate atomic and molecular structure in quantum chemistry.

\section{Charge Density in Quantum Chemistry}\label{QCsection}

To understand why Schr\"{o}dinger's role for amplitude-squared in determining charge density remains popular within quantum chemistry, we will examine two methods that are widely used to calculate ground state energies, electron densities, and arrangements of nuclei for atoms and molecules: the Hartree-Fock method and density functional theory.  Then, in section \ref{FEsection}, I will briefly discuss a few other places where the methods of calculation used in quantum chemistry provide evidence for Schr\"{o}dinger's electron charge density.

In both the Hartree-Fock method and density functional theory, we will see that the (approximate) ground state potential energy of an atom or molecule closely resembles the potential energy that would be calculated from Schr\"{o}dinger's electron charge density using purely classical physics.  This suggests that Schr\"{o}dinger's charge density accurately describes the true distribution of charge, even though the quantum expression for the energy associated with that charged matter is not exactly the same as the classical expression.

In classical electromagnetism, the mathematical representation of physical states tells you exactly where the electric charge is located.  In quantum mechanics, the use of wave functions leaves room for debate as to the placement of electric charge.  By comparing the classical and quantum expressions for potential energy, we can seek continuity between these theories and find evidence that will help us determine the fate of electric charge in the move from classical to quantum physics.

\subsection{The Hartree-Fock Method}\label{HFsection}

It is very difficult to accurately determine the ground state wave functions for atoms and molecules.  Approximations must be made.  The Hartree-Fock method has had a long history as one of the primary methods of approximation that quantum chemists use to calculate ground state wave functions.\footnote{The Hartree-Fock Method can also be used for excited states \cite[sec.\ 2.2.6]{szaboQC}, but we will not discuss that application here.}  When presenting this method, you commonly see authors employing Schr\"{o}dinger's role for amplitude-squared and speaking of electrons as clouds of charge.\footnote{See, for example, \cite[ch.\ 9]{slatervol1}; \cite[sec.\ 17.2]{slatervol2}; \cite[pg.\ 432]{blinder1965}; \cite[sec.\ 2.3.6]{szaboQC}; \cite[sec.\ 11.1]{levineQC}.}  In this section, we will see why that picture is appealing as a way of understanding the ground state energy in the Hartree-Fock Method.

Let us begin with a first step toward simplifying the problem of finding ground state wave functions for atoms and molecules.  In an atom, we can treat the nucleus as a point charge with its location fixed at the origin, generating a Coulomb potential that appears in the Hamiltonian which determines the ground state multi-electron wave function.  In a molecule, we can again treat the nuclei as point charges at particular locations.  In the Hartree-Fock method (and in density functional theory), the assumption of classical point nuclei at fixed locations can be relaxed by using the Born-Oppenheimer approximation to introduce a nuclear wave function.\footnote{See \cite[sec.\ 2.1.2]{szaboQC}; \cite[sec.\ 10.1]{mcquarrieQC}; \cite[ch.\ 8]{atkins2011}; \cite[sec.\ 13.1]{levineQC}.}  For our purposes here, we can keep things simple and stick with point nuclei.

To determine the locations of the nuclei in a molecule using the Hartree-Fock method, we can vary their arrangement, approximate the ground state electron wave function and the ground state energy of the molecule for each arrangement, and select the arrangement that minimizes ground state energy.\footnote{For certain combinations of atoms, there will be multiple local minima corresponding to different stable arrangements of the nuclei (structural isomers and stereoisomers).  See \cite{lowdin1989}; \cite[sec.\ 15.10]{levineQC}; \cite{franklinF}.}  Given certain locations for the nuclei, the task is then to determine the ground state wave function for the electrons.  Let us take the Hamiltonian\footnote{This Hamiltonian appears in \cite[pg.\ 345]{levineQC}.} in the time-independent Schr\"{o}dinger equation\footnote{Modifying the Hartree-Fock method, you can use the relativistic Dirac equation in place of the non-relativistic Schr\"{o}dinger equation.  With that modification, it is called the ``Dirac-Fock'' or ``Dirac-Hartree-Fock'' method \cite{desclaux2002}; \cite[pg.\ 581]{levineQC}.} to be:
\begin{equation}
\widehat{H}= \sum_{i=1}^{N}\left(\frac{-\hbar^2}{2 m} \nabla_i^2 \right)  + \sum_{k=1}^{M}\sum_{l>k}^{M} \left(\frac{q_k q_l}{|\vec{r}_k-\vec{r}_l|}\right) +  \sum_{i=1}^{N}\sum_{k=1}^{M}\left(\frac{-e q_k}{|\vec{x}_i-\vec{r}_k|}\right)  + \sum_{i=1}^{N}\sum_{j>i}^{N}\left(\frac{e^2}{|\vec{x}_i-\vec{x}_j|}\right)
\ ,
\label{hamiltonian}
\end{equation}
where the only interaction between particles that has been included is the electrostatic Coulomb interaction (written here in Gaussian cgs units).  In \eqref{hamiltonian}, the first term sums the kinetic energy operators for each electron, the second term sums Coulomb interactions between nuclei, the third term sums Coulomb interactions between electrons and nuclei, and the fourth term sums Coulomb interactions between electrons.  The $\vec{x}_i$ are coordinates for each of the $N$ electrons and the $\vec{r}_k$ are the given locations of the $M$ nuclei.  (For a single atom, $M=1$.)  The charge of nucleus $k$ is $q_k$.

One way to generate an antisymmetric $N$-electron wave function is to start with $N$ orthonormal single-electron wave functions\footnote{Usually, it is assumed that each of these spin orbitals can be written as the product of a single component complex-valued spatial wave function and a two-component spinor that has no spatial dependence.  In the restricted Hartree-Fock method (which can be used when there are an even number of electrons), one adds the assumption that electrons come in pairs having exactly the same spatial wave function and opposite spins (see \cite[ch.\ 3]{szaboQC}; \cite[pg.\ 12]{parryang}; \cite[sec.\ 7.17]{atkins2011}).}---written either as $\psi_i(\vec{x},s)$ or $\chi_i(\vec{x})$ and called ``spin orbitals'' or just ``orbitals''---and combine them in an antisymmetric product (the Slater determinant):
\begin{equation}
\psi(\vec{x}_1, s_1,\vec{x}_2, s_2, ...)=\frac{1}{\sqrt{N!}} \left| \begin{matrix}
\psi_1(\vec{x}_1, s_1) & \psi_2(\vec{x}_1, s_1) & \cdots & \psi_N(\vec{x}_1, s_1)\\
\psi_1(\vec{x}_2, s_2) & \psi_2(\vec{x}_2, s_2) & \cdots & \psi_N(\vec{x}_2, s_2)\\
\vdots & \vdots & \ddots & \vdots\\
\psi_1(\vec{x}_N, s_N) & \psi_2(\vec{x}_N, s_N) & \cdots & \psi_N(\vec{x}_N, s_N)
\end{matrix} \right|
\ .
\label{slaterdeterminant}
\end{equation}
Some antisymmetric wave functions can be written as Slater determinants, but others cannot.  In many cases, the lowest energy Slater determinant wave function is a good approximation to the actual ground state wave function.\footnote{For discussion of how good this approximation is, see \cite[ch.\ 18]{slatervol2}; \cite[pg.\ 164]{scerri1994}; \cite[sec.\ 11.1]{levineQC}.}  The Hartree-Fock method is aimed at estimating this lowest energy Slater determinant.  The difference between the energy of this Slater determinant and the true ground state wave function is called the ``correlation energy.''\footnote{The interpretation of correlation energy is discussed in \cite[sec.\ II.C]{lowdin1958}; \cite[pg.\ 436]{blinder1965}; \cite[sec.\ 9.8]{mcquarrieQC}.}  There are other ``post-Hartree-Fock'' methods that do not restrict themselves to Slater determinant wave functions, such as the configuration interaction method\footnote{This method is presented in \cite[ch.\ 4]{szaboQC}; \cite[sec.\ 9.8]{mcquarrieQC}; \cite[sec.\ 9.6]{atkins2011}; \cite[sec.\ 16.2]{levineQC}} (where one allows linear combinations of Slater determinants) and density functional theory (which will be discussed in section \ref{DFTsection}).  The assumption of a Slater determinant wave function allows us to write Schr\"{o}dinger's total electron charge density as a sum of contributions associated with each orbital (or one might say, with each electron) of $-e\chi_i^{\dagger}(\vec{x})\chi_i(\vec{x})$, but as that decomposition is not always available\footnote{See \cite{scerri2000, scerri2001, spence2001, matta2002}.} (and when it is, it is not necessarily unique\footnote{See footnote \ref{waterfootnote}.}) our focus in this article will be on the evidence regarding the status of Schr\"{o}dinger's total electron charge density for an atom or molecule (not individual electron charge densities).

The actual energy of the true ground state (its energy eigenvalue) is equal to the expectation value of the Hamiltonian for that state (because it is an energy eigenstate).  Other states will have higher expectation values.  We can approximate the ground state (and its energy) by searching for a Slater determinant wave function that minimizes the expectation value of the Hamiltonian,
\begin{equation}
\langle \widehat{H} \rangle = \sum_{s_1=1}^{2}\sum_{s_2=1}^{2}...\int{\psi^{*}(\vec{x}_1, s_1,\vec{x}_2, s_2, ...)\widehat{H}\psi(\vec{x}_1, s_1,\vec{x}_2, s_2, ...) \ d^3 x_1 d^3 x_2 ...}
\ .
\label{Hexpectation}
\end{equation}
There is much to be said about the techniques that are used to find the Slater determinant wave function that minimizes this expectation value, but let us put that aside and focus on understanding the ground state energy once we have it (or at least a good approximation to it).  The ground state energy \eqref{Hexpectation} will be the sum of a kinetic energy,
\begin{equation}
\sum_{i=1}^{N}\left(\frac{-\hbar^2}{2 m} \int{ \chi_i^{\dagger}(\vec{x})\nabla^2 \chi_i(\vec{x})\  d^3 x}\right)
\ ,
\label{electronkineticenergy}
\end{equation}
and a potential energy.  The potential energy can be written as the sum of five contributions.

The first contribution to the potential energy comes from the second term in \eqref{hamiltonian}.  It captures the potential energy of electrostatic Coulomb repulsion between the point nuclei and is independent of the electron wave function,
\begin{equation}
\sum_{k=1}^{M}\sum_{l>k}^{M} \left(\frac{q_k q_l}{|\vec{r}_k-\vec{r}_l|}\right)
\ .
\label{nucleusnucleusenergy}
\end{equation}

The second contribution to the potential energy comes from the third term in \eqref{hamiltonian}.  To calculate this contribution, we can sum the expectation values for the energy of Coulomb attraction between the electron and the nuclei for each electron orbital,
\begin{equation}
\sum_{i=1}^{N}\sum_{k=1}^{M}\left(-e q_k \int{ \frac{\chi_i^{\dagger}(\vec{x})\chi_i(\vec{x})}{|\vec{x}-\vec{r}_k|}\  d^3 x}\right)
\ .
\label{electronnucleusenergy}
\end{equation}
This contribution to the energy is exactly what we would expect classically\footnote{In classical electrostatics, the density of potential energy for a distribution of charge $\rho^q(\vec{x})$ in a potential $V(\vec{x})$ can be written as $\frac{1}{2}\rho^q(\vec{x}) V(\vec{x})$ \cite[sec.\ 1.11]{jackson}; \cite[sec.\ 2.4]{griffiths}.  Because there will be an equal contribution to the potential energy from the nuclei in the potential sourced by the electron charge density $\rho^q(\vec{x})$, \eqref{electronnucleusenergy2} does not contain the factor of $\frac{1}{2}$.  Moving from electrostatics to electromagnetism, this potential energy of charged matter is replaced by electromagnetic field energy \cite[ch.\ 5]{lange}; \cite[sec.\ 2.4.4]{griffiths}.  However, for our purposes here (where we are only concerned with electrostatic Coulomb attraction and repulsion), it will be convenient to treat this kind of energy as potential energy possessed by matter.} as the potential energy associated with a distribution of charge
\begin{equation}
\rho^q(\vec{x})=\sum_{i=1}^{N}-e\chi_i^{\dagger}(\vec{x})\chi_i(\vec{x})
\ ,
\label{chargedensity1}
\end{equation}
interacting electrostatically with a collection of point charges $q_k$ at locations $\vec{r}_k$, producing an electric potential
\begin{equation}
V(\vec{x})=\sum_{k=1}^{M}\frac{q_k}{|\vec{x}-\vec{r}_k|} 
\ .
\label{potentialfromnuclei}
\end{equation}
In this notation, \eqref{electronnucleusenergy} can be written as
\begin{equation}
\int{\rho^q(\vec{x}) V(\vec{x}) \  d^3 x}
\ .
\label{electronnucleusenergy2}
\end{equation}
Because \eqref{chargedensity1} is the charge density that would be calculated from the full wave function using Schr\"{o}dinger's proposal from the previous section \eqref{totalchargedensitypauli},\footnote{See \cite[pg.\ 9]{bader1990}; \cite[problem 16.28]{levineQC}.}  the presence of this contribution to the potential energy suggests that electron charge really is spread out in the way Schr\"{o}dinger proposed.\footnote{Note that the availability of this interpretation is independent of the Hartree-Fock approximation.  For any wave function (Slater determinant or not), the contribution to the expectation value of the energy from the third term in \eqref{hamiltonian} is equal to the classical electrostatic energy of the Schr\"{o}dinger charge distribution \eqref{totalchargedensitypauli} derived from that wave function interacting with the point charge nuclei.}  In the next contribution, we will see further corroboration of Schr\"{o}dinger's proposal.

There is only one term remaining in the Hamiltonian \eqref{hamiltonian} (the electron-electron interaction term), but we will separate the potential energy associated with this term into three pieces.\footnote{The derivation of \eqref{electronelectronrepulsionenergy}--\eqref{electronelectronexchangeenergy} from the electron-electron interaction term in \eqref{hamiltonian} appears in \cite{blinder1965}.}  First among these, we have our third contribution to the total potential energy,
\begin{equation}
\sum_{i=1}^{N}\sum_{j \geq i}^{N}\left(e^2\int{\frac{\chi_i^{\dagger}(\vec{x})\chi_i(\vec{x})\chi_j^{\dagger}(\vec{x}')\chi_j(\vec{x}')}{|\vec{x}-\vec{x}'|}\  d^3 x d^3 x'}\right)
\ .
\label{electronelectronrepulsionenergy}
\end{equation}
This is equal to the classical electrostatic energy of the total charge distribution \eqref{chargedensity1} derived from the multi-electron wave function interacting with itself via Coulomb repulsion, and can be written as
\begin{equation}
\frac{1}{2}\int{\frac{\rho^q(\vec{x}) \rho^q(\vec{x}')}{|\vec{x}-\vec{x}'|}\  d^3 x d^3 x'}
\ .
\label{electronelectronrepulsionenergy2}
\end{equation}
As with \eqref{electronnucleusenergy}, the presence of this contribution to the energy supports Schr\"{o}dinger's idea that the charge of the electrons is spread out as described in \eqref{totalchargedensitypauli}.  We can think of the remaining two contributions as quantum corrections that make the actual energy lower than the classical energy associated with this charge distribution.

The fourth contribution to the potential energy is
\begin{equation}
-\sum_{i=1}^{N}\left(e^2\int{\frac{\chi_i^{\dagger}(\vec{x})\chi_i(\vec{x})\chi_i^{\dagger}(\vec{x}')\chi_i(\vec{x}')}{|\vec{x}-\vec{x}'|}\  d^3 x d^3 x'}\right)
\ .
\label{electronselfenergy}
\end{equation}
Thinking of the electron in each orbital as spread out with charge density $-e\chi_i^{\dagger}(\vec{x})\chi_i(\vec{x})$, \eqref{electronselfenergy} subtracts out the energy of self-repulsion that would arise from each electron's charge distribution interacting with itself.\footnote{By treating the nuclei as point charges and only considering contributions to the energy from pairs of distinct nuclei, we have also left out any contributions from nuclear self-repulsion.}  For a single-electron atom,  \eqref{electronelectronrepulsionenergy} and \eqref{electronselfenergy} will cancel so that there is no net contribution to the potential energy from electrostatic repulsion.  However, for multi-electron atoms and molecules \eqref{electronelectronrepulsionenergy} will be larger than \eqref{electronselfenergy}.  These two contributions can be combined into a single ``Coulomb integral'' that gives classical energy of electrostatic repulsion between every pair of distinct electron orbitals,
\begin{equation}
\sum_{i=1}^{N}\sum_{j > i}^{N}\left(e^2\int{\frac{\chi_i^{\dagger}(\vec{x})\chi_i(\vec{x})\chi_j^{\dagger}(\vec{x}')\chi_j(\vec{x}')}{|\vec{x}-\vec{x}'|}\  d^3 x d^3 x'}\right)
\ ,
\label{coulombintegral}
\end{equation}
which differs from \eqref{electronelectronrepulsionenergy} only in that the $\geq$ in the second sum is replaced by a $>$.\footnote{Schr\"{o}dinger \cite[sec.\ 3]{schrodinger1928report} applied his charge density role for amplitude-squared to the interpretation of such Coulomb integrals when analyzing multi-electron atoms.  In the included discussion with Born, Heisenberg, and Fowler, Schr\"{o}dinger briefly discusses Hartree's method (a precursor to Hartree-Fock that leaves out the exchange integrals) in order to show that his ``hope of achieving a three-dimensional conception [a physical understanding of what is happening in three-dimensional space] is not quite utopian,'' though he notes the remaining challenge of incorporating and interpreting exchange terms \cite[pg.\ 428--429]{bacciagaluppi2009}.}

The fifth contribution to the potential energy is called the ``exchange integral,''
\begin{equation}
-\sum_{i=1}^{N}\sum_{j>i}^{N}\left(e^2\int{\frac{\chi_i^{\dagger}(\vec{x})\chi_j(\vec{x})\chi_j^{\dagger}(\vec{x}')\chi_i(\vec{x}')}{|\vec{x}-\vec{x}'|}\  d^3 x d^3 x'}\right)
\ .
\label{electronelectronexchangeenergy}
\end{equation}
Sometimes \eqref{electronselfenergy} and \eqref{electronelectronexchangeenergy} are combined into a single expression, replacing the $>$ in the second sum above by a $\geq$.\footnote{This way of rearranging terms is mentioned in \cite[pg.\ 436]{blinder1965}; \cite[pg.\ 5048]{perdew1981}; \cite[pg.\ 7]{parryang}.}  The exchange integral \eqref{electronelectronexchangeenergy} does not have a straightforward interpretation as the energy associated with a classical spread-out charge distribution (and was not included in the original Hartree method, which predates the Hartree-Fock method presented here).  The exchange integral arises because the wave function is an antisymmetric product of single-electron orbitals (a Slater determinant), not a simple product of these orbitals.

Using the Hartree-Fock method, we have written an approximation to the ground state potential energy of an atom or molecule as the sum of five contributions:
\begin{enumerate}
\setlength\itemsep{0 pt}
\item Nucleus-Nucleus Repulsion \eqref{nucleusnucleusenergy}
\item Minus Electron-Nucleus Attraction \eqref{electronnucleusenergy}
\item Plus Electron Repulsion \eqref{electronelectronrepulsionenergy}
\item Minus Electron Self-Repulsion \eqref{electronselfenergy}
\item Minus Exchange Energy \eqref{electronelectronexchangeenergy}
\end{enumerate}
We have seen that the first three contributions can be interpreted as giving the classical electrostatic potential energy associated with Schr\"{o}dinger's electron charge distribution interacting with the point nuclei (including repulsion between the nuclei and within the electron charge distribution) and the final two contributions as describing quantum corrections to the potential energy.  I think this gives us strong reason to take Schr\"{o}dinger's proposal seriously.  One might respond that Schr\"{o}dinger's charge distribution is inadequate because it fails to give a classical explanation of all five contributions to the potential energy.  There are ways the details might be filled in so that it succeeds in doing so (to be discussed in section \ref{BMsection}), but I don't see why it must.  It is not surprising that the expression for energy in quantum physics is different from classical electromagnetism.  The fact that the classical energy of Schr\"{o}dinger's charge distribution appears as a contribution to the total energy makes it appealing to say that we have familiar stuff (spread-out matter with charge) obeying new rules (quantum physics).

Having explained how the Hartree-Fock ground state energy can be interpreted in terms of a spread-out electron charge density, one might wonder whether the same expressions can alternatively be understood in terms of point electrons.  Let us look first at the contribution to the energy from electron-nucleus interaction \eqref{electronnucleusenergy}.  As a contribution to the expectation value of energy, \eqref{electronnucleusenergy} can be interpreted as a probabilistically weighted average of the potential energies of pairs of point electrons and point nuclei---considering different possible locations for the electrons, with the probability density for each electron being given by $\chi_i^{\dagger}(\vec{x})\chi_i(\vec{x})$.  However, as one of the contributions to the actual ground state energy at a particular moment, \eqref{electronnucleusenergy} has no natural interpretation in terms of point charges.  Classically, this energy should instead be
\begin{equation}
\sum_{i=1}^{N}\sum_{k=1}^{M}\left(\frac{-e q_k}{|\vec{x}_i-\vec{r}_k|}\right)
\ ,
\label{pointelectronnucleusenergy}
\end{equation}
where $\vec{x}_i$ are the actual locations of the point electrons.  Similarly, the contribution to the expectation value of the energy from the fourth term in \eqref{hamiltonian} can be interpreted as a weighted average of the potential energies associated with electron-electron repulsion, with the probability density for a particular electron configuration given by the square of the full wave function.  In the decomposition of this contribution, the Coulomb integral \eqref{coulombintegral} can be interpreted as a weighted average of the potential energies associated electron-electron interaction assuming that each electron has an independent probability distribution $\chi_i^{\dagger}(\vec{x})\chi_i(\vec{x})$ and the exchange integral can be interpreted as a correction that accounts for the fact that the probabilities for electrons being in certain locations are not independent.  The antisymmetrized wave function generated by taking a Slater determinant assigns low probabilities to two electrons with the same spin being close to one another.  This is sometimes explained by saying that, as a consequence of the Pauli exclusion principle, each electron is surrounded by a ``Fermi hole'' where other electrons are unlikely to be found \cite[pg.\ 218]{lowdin1958}; \cite[sec.\ 17.2]{slatervol2}.

For each of the terms, the point charge picture of the electron offers a way to understand the expectation values for different contributions to the potential energy.  But, what about the actual values of these contributions to the potential energy of an atom or molecule?  On the face of it, the point electron picture seems to get these wrong.  There are a number of possible responses available to the proponent of point electrons.  One could just say that the potential energy of an atom or molecule is determined by the wave function, not the actual positions of the point electrons (though this would mean that the point electrons do not act very much like point charges).  There is a more ambitious strategy (that was mentioned in section \ref{PCsection} and will be returned to in section \ref{BMsection}): perhaps electrons move rapidly throughout the electron cloud of an atom or molecule so that Schr\"{o}dinger's charge density is the time-averaged charge density (over a short period of time) and what we ordinarily call ``the potential energy'' of an atom or molecule is really the time-averaged potential energy.  Finally, there is a radical strategy where we regard the various different possible locations of electrons---that are integrated over in calculating expectation values---as actual locations of electrons in different universes (an idea that will be elaborated on in section \ref{BMsection}).

\begin{figure}[htb]
\center{\includegraphics[width=5 cm]{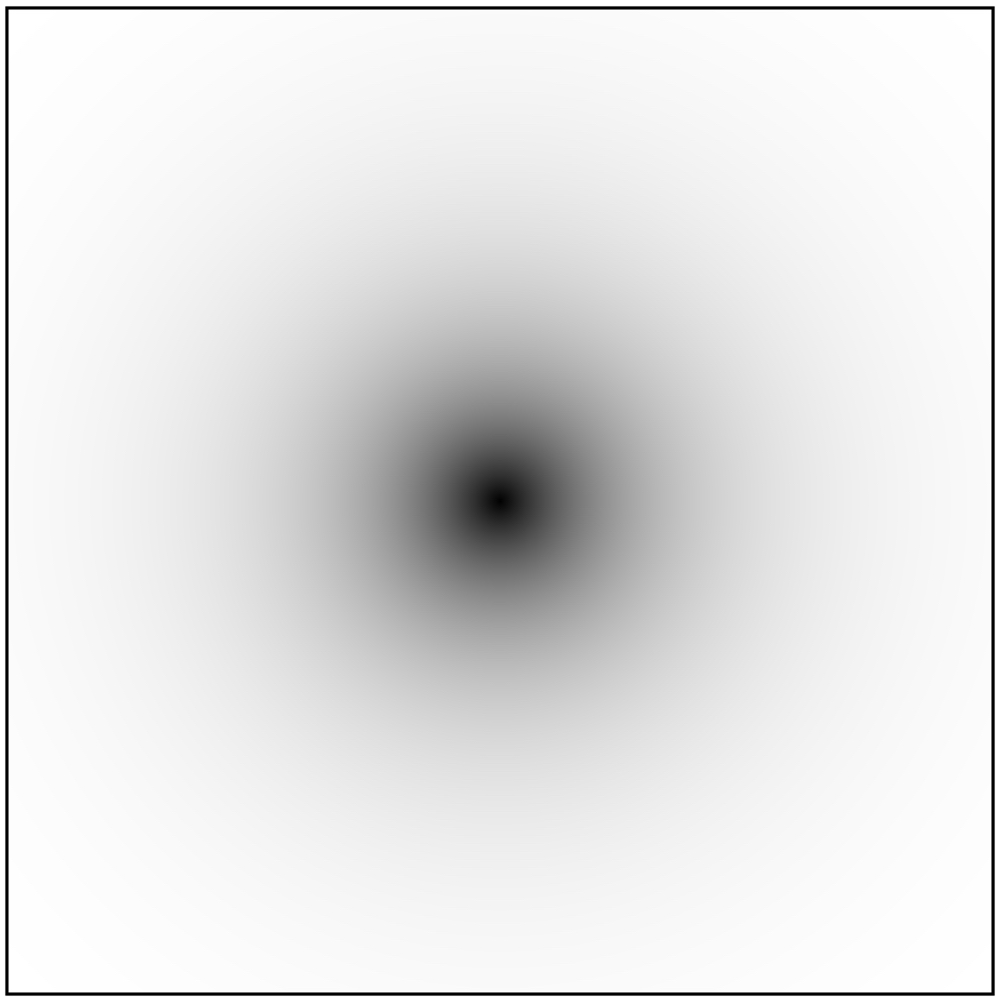}}
\caption{This figure shows the electron charge density for the helium atom (the electron cloud), peaked in the center at the nucleus.  Because the two orbitals \eqref{twoorbitals} have the same charge density, this image would also be an accurate representation of the charge density associated with either orbital.}
\label{helium}
\end{figure}

To better understand how the Hartree-Fock method works and how it supports Schr\"{o}dinger's role for amplitude-squared, let us consider the ground states of the helium atom and the water molecule.  First, the helium atom.  Because the helium atom has two electrons with their spins paired, we can take the orbitals to have the general form
\begin{align}
\chi_1(\vec{x})&=f(\vec{x})\left(
\begin{matrix}
1 \\
0
\end{matrix}
\right)
\nonumber
\\
\chi_2(\vec{x})&=f(\vec{x})\left(
\begin{matrix}
0 \\
1
\end{matrix}
\right)
\ .
\label{twoorbitals}
\end{align}
The task is then to find $f(\vec{x})$, the ``spatial wave function.''  As an educated guess,\footnote{See \cite[sec.\ 9.4 and 11.1]{levineQC}.} we can assume that this function is the sum of multiple functions that all resemble the spatial wave function for the 1s orbital of the hydrogen atom,
\begin{equation}
f(\vec{x})=\frac{1}{\sqrt{\pi}} \sum_{i=1}^{5} c_i \left(\frac{\zeta_i}{a_0}\right)^{3/2}e^{-\zeta_i |\vec{x}| / a_0}\ ,
\label{fexpansion}
\end{equation}
where here we have included just five terms in the sum.  In \eqref{fexpansion}, $a_0$ is the Bohr radius, $\frac{\hbar^2}{m e^2}$.  We can calculate the constants $c_i$ and $\zeta_i$ by minimizing the expectation value of the Hamiltonian \eqref{Hexpectation}.  Levine \cite[sec.\ 11.1]{levineQC} discusses this example and reports values for these constants from \cite{clementi1974}: $c_1=0.76838$, $c_2=0.22346$, $c_3=0.04082$, $c_4=-0.00994$, $c_5=0.00230$, $\zeta_1=1.41714$, $\zeta_2=2.37682$, $\zeta_3=4.39628$, $\zeta_4=6.52699$, $\zeta_5=7.94252$.\footnote{McQuarrie \cite[pg.\ 482--489]{mcquarrieQC} reviews in detail a similar Hartree-Fock calculation of the ground state of the helium atom, summing over just two terms instead of the five in \eqref{fexpansion}.}  Schr\"{o}dinger's charge density \eqref{chargedensity1} for this ground state wave function is depicted in figure \ref{helium}.

For the approximate ground state wave function described above, we can break down the total ground state energy into various contributions.  In molecules, there will be an energy of nucleus-nucleus repulsion \eqref{nucleusnucleusenergy}.  But, for the helium atom this contribution is zero as there is only a single nucleus.  There is a negative energy associated with electron-nucleus attraction of 183.7 eV, which can be calculated from \eqref{slaterdeterminant}, \eqref{electronnucleusenergy}, \eqref{twoorbitals}, and \eqref{fexpansion}.  This is exactly the classical electrostatic energy associated with the interaction between a positive point charge $2e$ (the nucleus) and Schr\"{o}dinger's negative electron charge density \eqref{chargedensity1}.  Classically, the energy of this charge density interacting with itself would be 83.7 eV.  This is the contribution to the total energy from electron repulsion \eqref{electronelectronrepulsionenergy}.  That includes contributions to the energy from the self-repulsion within each electron's cloud and the interaction between the two electron clouds.  To remove the self-repulsion terms, we must subtract out a self-repulsion energy of 27.9 eV for each electron \eqref{electronselfenergy}.  There is no exchange energy \eqref{electronelectronexchangeenergy} for electrons in the Helium atom because the orbitals in \eqref{twoorbitals} are orthogonal (the electrons have opposite spin).  Finally, there is a kinetic energy \eqref{electronkineticenergy} of 77.9 eV.  Putting all of this together yields a total ground state energy for the helium atom of
\begin{equation}
-183.7\mbox{ eV}+83.7\mbox{ eV}-(2 \times 27.9\mbox{ eV})+77.9\mbox{ eV}=-77.9\mbox{ eV}
\ .
\end{equation}

In the above calculation, subtracting out the energies of self-repulsion significantly changes the ground state energy.  In this way, Schr\"{o}dinger's charge density does not act like an ordinary classical charge density.  But, it is not so surprising that this correction would be missed in a classical theory of electromagnetic interactions at macroscopic scales.  When you calculate the electrostatic potential energy of, say, a charged Van de Graaff generator, you are primarily integrating over interactions between distinct particles at distant locations.  Whether you include self-repulsion or not makes no difference.  It is only when we study subatomic particles that we can determine whether self-repulsion makes a contribution to the total energy and, as it turns out, it doesn't.  The fact that we have to subtract out self-repulsion does not mean that Schr\"{o}dinger's charge density is not a real charge density.  It just means that electric charge in quantum mechanics acts a bit differently from electric charge in classical electromagnetism.  The presence of exchange energy \eqref{electronelectronexchangeenergy} can also be interpreted as revealing a difference in the properties of classical and quantum charged matter.  Despite these differences, Schr\"{o}dinger's density of charge in quantum mechanics acts enough like charge density in classical electromagnetism that it can be recognized as a true density of charge (obeying the laws of a new physical theory).

\begin{figure}[p]
\center{\includegraphics[width=8 cm]{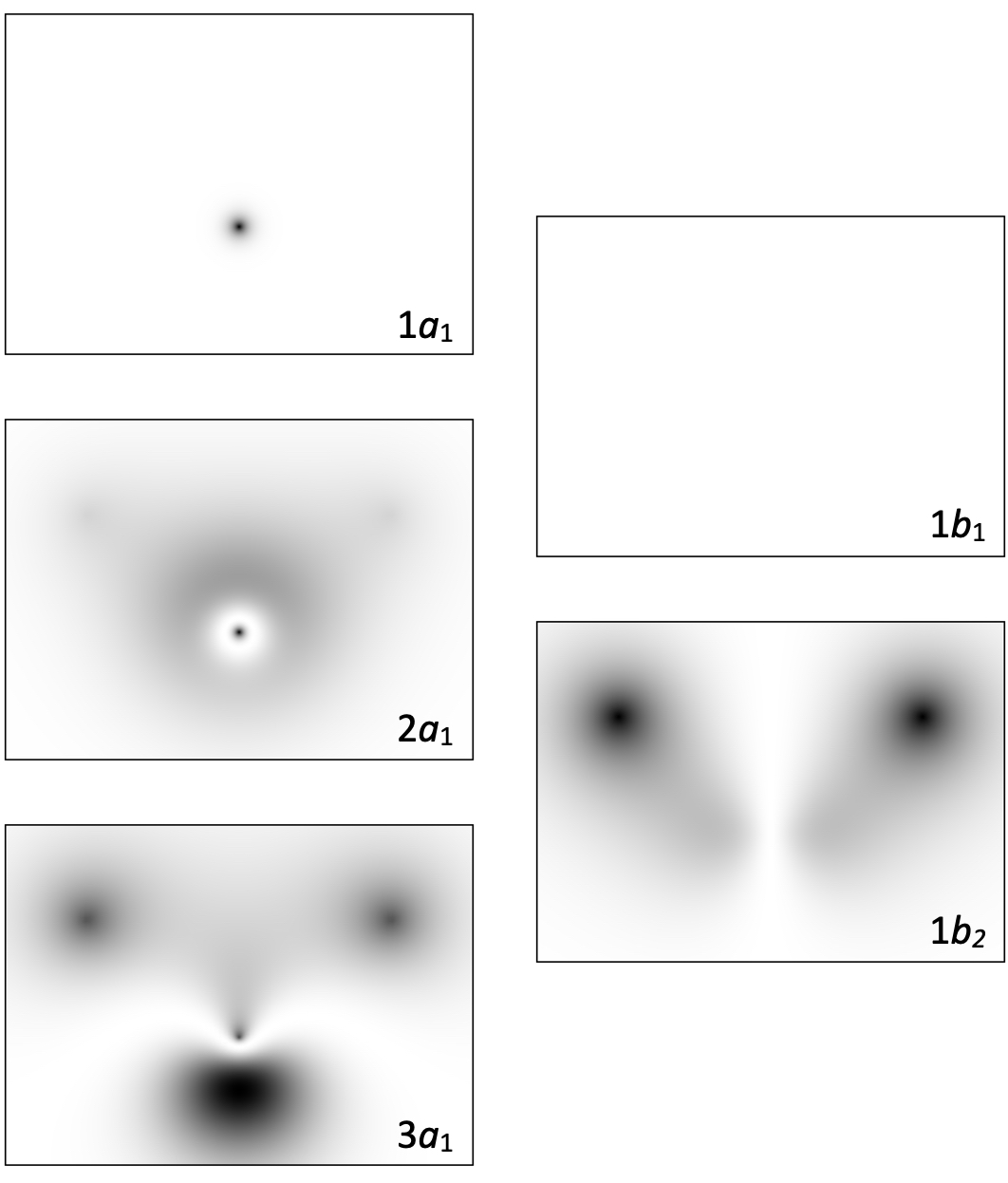}}
\caption{These density plots depict the charge densities, $-e\chi_i^{\dagger}(\vec{x})\chi_i(\vec{x})$, of 10 electron orbitals for a ground state water molecule, calculated via the Hartree-Fock method.  There are only five images because the orbitals come in pairs with opposite spin and identical charge density (and thus a single plot suffices to represent each pair of orbitals).  These plots show the charge density in the plane picked out by the nuclei (which happens to be zero for the pair of orbitals labeled ``$1b_1$'').  The two hydrogen nuclei are near the upper corners and the oxygen nucleus is just below the center in each image.  These density plots have been generated from the molecular orbitals found using the Hartree-Fock method in \cite{aung1968,pitzer1970}, as reported by Levine \cite[sec.\ 15.5]{levineQC}.  Levine \cite[pg.\ 452]{levineQC} describes the $2a_1$ and $1b_2$ orbitals as bonding orbitals because between the nuclei there is what he calls an ``electron probability-density buildup'' or (slipping without comment from Born's role to Schr\"{o}dinger's) an ``electron charge buildup.''  This buildup of charge is visible in the above plots.  Although the Hartree-Fock method can be used to determine the arrangement of nuclei in the molecule, the orbitals in these images were found using experimental values for the bond angle and bond lengths.  These density plots can be compared to the depictions of similar molecular orbitals for the H$_2$O molecule in \cite{dunning1972}.}
\label{waterorbitals}
\end{figure}

\begin{figure}[htb]
\center{\includegraphics[width=8 cm]{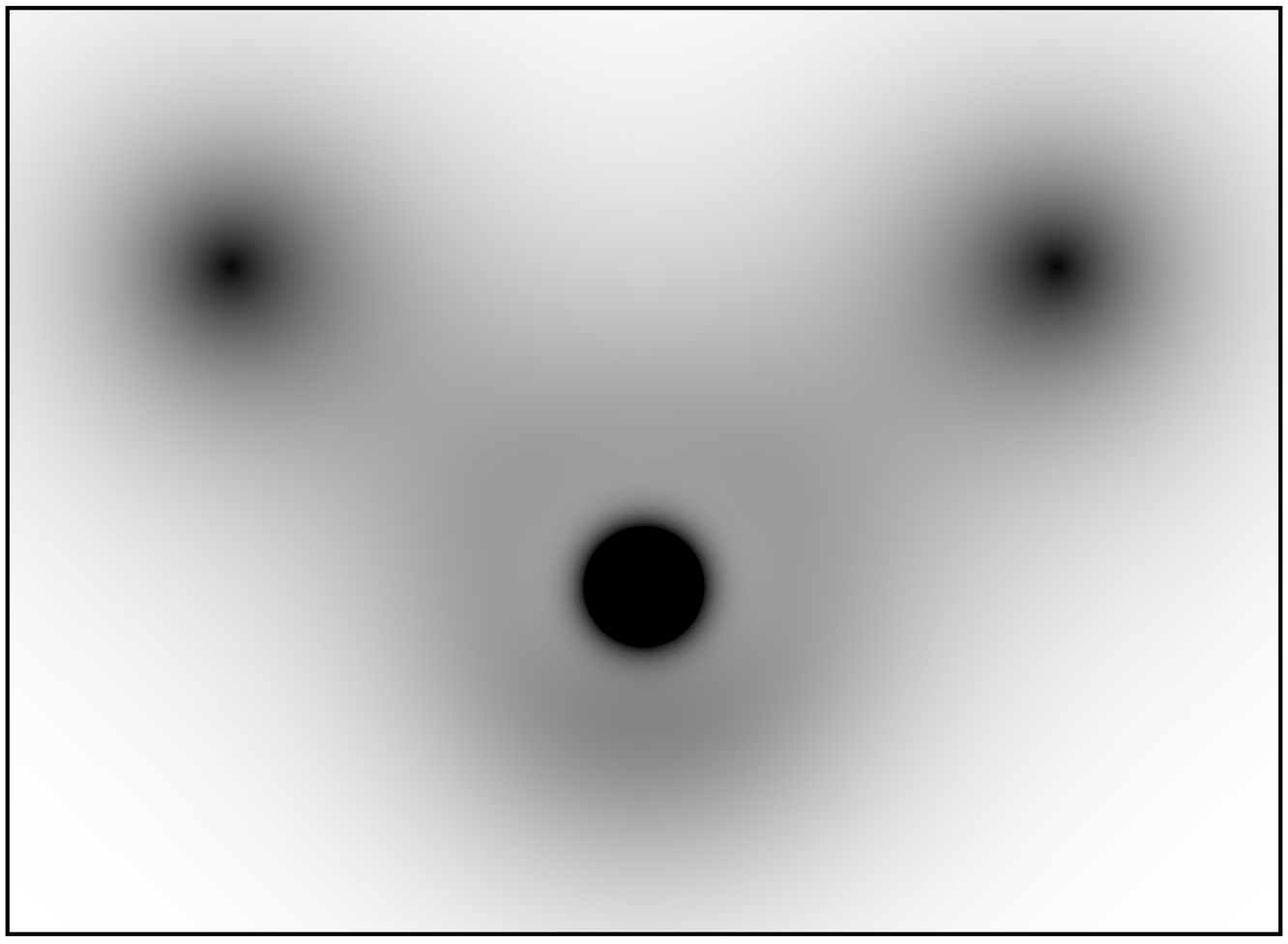}}
\caption{This figure depicts the total electron charge density for the water molecule \eqref{chargedensity1}, found by summing the charge densities for each orbital depicted in figure \ref{waterorbitals}.  In this density plot, darker gray indicates a higher density of charge and black indicates a charge density that is at or above a certain threshold.  In the circular black region around the oxygen nucleus, the density exceeds this threshold (and thus variation in density within that region is not shown).  Introducing this threshold charge density allows us to focus on areas where the charge density is weak and thereby helps to highlight the concentration of charge density along the bond paths between the hydrogen nuclei and the oxygen nucleus.}
\label{water}
\end{figure}

We just saw that for the helium atom the Hartree-Fock method can be used to find the ground state electron wave function and the ground state energy.  For molecules, the Hartree-Fock method allows us to determine not only the electron wave function, but also the arrangement of nuclei.  Applying the Hartree-Fock method to the water molecule (H$_2$O), one can calculate the distances between the nuclei and the bond angle\footnote{Calculations of the bond angle of H$_2$O have been discussed in the literature on scientific realism as a point of agreement between the different interpretations of quantum mechanics \cite[pg.\ S309]{cordero2001}; \cite[sec.\ 4.5.2]{callender2020}.} \cite[pg.\ 202]{szaboQC}; \cite[sec.\ 15.5]{levineQC}.  This is done by varying the locations of the nuclei and minimizing the expectation value of the Hamiltonian \eqref{Hexpectation} for each arrangement (then selecting the arrangement with the lowest value).  In the end, we arrive at a ground state energy, locations for the three nuclei, and a set of 10 electron orbitals (figure \ref{waterorbitals}).\footnote{As Levine \cite[sec.\ 15.8]{levineQC} explains, the Hartree-Fock method does not (and cannot) yield a unique set of orbitals.  There are alternative sets of orbitals that lead to the same multi-electron wave function when you take the Slater determinant.\label{waterfootnote}}  These orbitals together form the multi-electron wave function when we take the Slater determinant.  From this wave function, we can calculate the electron charge density (figure \ref{water}).

\subsection{Density Functional Theory}\label{DFTsection}

The Hartree-Fock method is an old and venerable standby for calculating the ground state energies and electron structures of atoms and molecules.  Density functional theory is a newer alternative that has become very popular.\footnote{See \cite[pg.\ 69]{argaman2000}; \cite{becke2014}.}  Let us begin by dissecting the name.  Despite the word ``theory,'' density functional theory is not a new physical theory.  Instead, like the Hartree-Fock method, it is a technique for approximating various properties of atoms and molecules.  Unlike the Hartree-Fock method, density functional theory does not attempt to calculate ground state multi-electron wave functions but instead focuses on the ground state ``electron density''---hence the ``density'' in density functional theory.  What kind of density is the electron density?  When presenting density functional theory, some authors describe the electron density as an expected number density \eqref{numdensitypauli} and others as a charge density \eqref{totalchargedensitypauli}.  Here, we will treat it as a density of charge.

In their quantum chemistry textbooks, McQuarrie \cite[pg.\ 649--650]{mcquarrieQC}, Atkins and Friedman \cite[pg.\ 317]{atkins2011}, and Levine \cite[sec.\ 16.5]{levineQC} introduce the density of density functional theory as an ``electron probability density,'' though it would be more accurate to call it an ``expected number density.''\footnote{Although Levine \cite[sec.\ 16.5]{levineQC}  calls the electron density an ``electron probability density,'' he does not mean by this that integrating the electron density \eqref{numdensitypauli} over any spatial region gives the probability of finding at least one electron in that region.  For sufficiently small regions, the probability of finding more than one electron in the region is negligible and the integrated electron density over that region can be interpreted as giving the probability of finding one (or at least one) electron in that region.  However, for large regions the possibility of finding multiple electrons cannot be neglected.  The integral of the electron density \eqref{numdensitypauli} over all space is $N$ (the total number of electrons) not $1$ (the probability of finding at least one electron if you look everywhere).  Thus, the electron density may be interpreted as an expected number density (as was explained in section \ref{PCsection}) but, speaking precisely, it cannot be interpreted as a probability density.}  Parr and Yang \cite[pg.\ 14]{parryang}, in an early and influential textbook on density functional theory, initially describe the electron density as a number density, giving no indication that it is only an expected number density and thus appearing to interpret it as an actual number density.  If it is an actual number density, then (as was discussed in section \ref{PCsection}) we cannot be thinking of electrons as point particles but must instead be thinking of electrons and their charge as spread out (with the density of charge proportional to the number density).  Taking this view, Bader \cite{bader2010} understands the electron density to be a density of charge.  Many authors jump into presentations of density functional theory without explicitly specifying the type of density that is being studied \cite{seminario1995, martin2004, engel2011, becke2014}.


Density functional theory takes as its jumping off point the Hohenberg-Kohn theorem, showing that the ground state antisymmetric multi-electron wave function for an atom or molecule is uniquely determined by the ground state electron charge density.  Generalizing from \eqref{hamiltonian}, we can write the Hamiltonian for a collection of electrons in the presence of some fixed external distribution of charge that generates an electrostatic scalar potential $V(\vec{x})$ as
\begin{equation}
\widehat{H}= \sum_{i=1}^{N}\left(\frac{-\hbar^2}{2 m} \nabla_i^2 \right)  + E_{nn}[V] + \sum_{i=1}^{N}(-e)V(\vec{x}_i) + \sum_{i=1}^{N}\sum_{j>i}^{N}\left(\frac{e^2}{|\vec{x}_i-\vec{x}_j|}\right)
\ ,
\label{hamiltonian2}
\end{equation}
where $E_{nn}[V]$ is the potential energy of interaction among the charges producing the external potential (which in this context would be the atomic nuclei).  The Hohenberg-Kohn theorem tells us that for a given electron charge density $\rho^q(\vec{x})$ there is (up to a global additive constant) a unique potential $V^*(\vec{x})$ that can be plugged into \eqref{hamiltonian2} so that the ground state wave function for this Hamiltonian has the charge density $\rho^q(\vec{x})$.  The potential in the Hamiltonian and the ground state wave function are fixed once we have specified the ground state charge density.  As a consequence of this theorem, any property of the ground state that is fixed by the multi-electron wave function (like the total energy or the kinetic energy) can alternatively be fixed by specifying the charge density.

Recall that for a ground state atom or molecule, the actual energy is equal to the expectation value of the Hamiltonian \eqref{Hexpectation}.  In terms of the charge density, we can thus write the ground state energy as\footnote{This way of writing the energy functional matches \cite[eq.\ 3.2.3]{parryang}; \cite[eq.\ 9.36]{atkins2011}; \cite[eq.\ 16.36]{levineQC}, though here I have also included the potential energy associated with repulsion between the nuclei to parallel the presentation of the Hartree-Fock method in section \ref{HFsection} (as in \cite[eq.\ 6.12]{martin2004}).}
\begin{equation}
E[\rho^q]=\overline{T}[\rho^q]+E_{nn}[V^*]+\int{\rho^q(\vec{x}) V^*(\vec{x}) \  d^3 x}+\overline{V}_{ee}[\rho^q]
\ ,
\label{energyfunctional1}
\end{equation}
where $V^*$ is the potential picked out by $\rho^q$ via the Hohenberg-Kohn theorem.  The expression for the energy in \eqref{energyfunctional1} is called an energy ``functional'' because the total energy is written as a function of the electron charge density, which is itself a function of position.  The fact that properties like total energy can be written as functionals of the density explains the presence of ``density functional'' in the name ``density functional theory.''

The first term in \eqref{energyfunctional1}, $\overline{T}[\rho^q]$, is an as-yet-unspecified functional giving the kinetic energy of the electrons.  The second term is the electrostatic potential energy of repulsion between the external charges.  For point nuclei, this is given by \eqref{nucleusnucleusenergy}.  The third term is the classical potential energy associated with the interaction between the electron charge distribution $\rho^q(\vec{x})$ and the external charge distribution that generates the potential $V(\vec{x})$, \eqref{electronnucleusenergy2}.  The fourth term, $\overline{V}_{ee}[\rho^q]$, is an as-yet-unspecified functional giving the energy of electron-electron repulsion.

We can extend the energy functional in \eqref{energyfunctional1} so that it takes as input both an electron charge density $\rho^q(\vec{x})$ and a potential $V(\vec{x})$.\footnote{This extension is explained clearly in \cite[pg.\ 5384]{baerends1997}.}  To simplify our analysis, let us suppose that the potential $V$ arises from a collection of point nuclei with locations $\vec{r}_k$ and charges $q_k$ \eqref{potentialfromnuclei}.  The extended energy functional
\begin{equation}
E[\rho^q,V]=\overline{T}[\rho^q]+\sum_{k=1}^{M}\sum_{l>k}^{M} \left(\frac{q_k q_l}{|\vec{r}_k-\vec{r}_l|}\right)+\int{\rho^q(\vec{x}) V(\vec{x}) \  d^3 x}+\overline{V}_{ee}[\rho^q]
\label{energyfunctional2}
\end{equation}
returns the expectation value of the Hamiltonian in \eqref{hamiltonian} with the inputted potential $V(\vec{x})$ (and the corresponding nuclei locations and charges), where this expectation value is evaluated for the multi-electron wave function that the inputted $\rho^q(\vec{x})$ picks out as the ground state for some potential $V^*(\vec{x})$ (by the Hohenberg-Kohn theorem).  The same functional $\overline{T}[\rho^q]$ that gave the actual kinetic energy in \eqref{energyfunctional1} gives the expected kinetic energy in \eqref{energyfunctional2}.  Similarly, $\overline{V}_{ee}[\rho^q]$ in \eqref{energyfunctional2} gives the expected energy of electron-electron repulsion.  If $\rho^q(\vec{x})$ is the ground state charge density for $V(\vec{x})$, then $V(\vec{x})=V^*(\vec{x})$ and this energy functional returns the actual ground state energy ($E[\rho^q,V]=E[\rho^q]$).  The Hohenberg-Kohn variational theorem (a.k.a.\ the second Hohenberg-Kohn theorem) says that if we hold $V(\vec{x})$ fixed and vary $\rho^q(\vec{x})$, the minimum value for $E[\rho^q,V]$ will be attained when $\rho^q(\vec{x})$ is the ground state charge density for $V(\vec{x})$.

As with the Hartree-Fock method, the project here is one of minimization.  We can consider candidate locations for the nuclei and charge densities for the electrons, searching for the combination that minimizes the energy functional \eqref{energyfunctional2}.  Succeeding in this endeavor would yield the ground state energy, the locations of nuclei, and the electron charge density.  In principle, these would be approximate only because the nuclei are being modeled as classical point charges and the quantum theory we are using is a non-relativistic approximation to deeper physics.  However, in practice further approximation is necessary because we do not have exact expressions for $\overline{T}[\rho^q]$ and $\overline{V}_{ee}[\rho^q]$.  There are a variety of techniques that are used to approximate these quantities.

Before we go further, note that the output of this method is different from that of the Hartree-Fock method.  With the Hartree-Fock method, we got an approximate ground state energy and an approximate multi-electron wave function---learning about ``electron structure'' at the level of the quantum wave function.  With density functional theory, we get an approximate ground state energy and an approximate electron charge density---learning about electron structure only at the coarse-grained level of charge density and not at the more fine-grained level of the multi-electron wave function.  This charge density selects a ground state wave function, but density functional theory allows us to bypass the project of finding this wave function.  In contrast to the Hartree-Fock method, we do not require that the selected wave function be expressible as a Slater determinant of electron orbitals.  The above differences suggest a potential advantage of density functional theory: instead of varying nuclei locations and a multi-electron wave function on high-dimensional configuration space to minimize \eqref{Hexpectation}, we can, in principle, simply vary nuclei locations and an electron charge density on three-dimensional space to minimize \eqref{energyfunctional2}.  The decrease in dimensionality simplifies the search.

To move forward without knowing $\overline{T}[\rho^q]$ and $\overline{V}_{ee}[\rho^q]$, we can rewrite \eqref{energyfunctional2} in a widely used form due to Kohn and Sham \cite{kohn1965},\footnote{See \cite[eq.\ 7.1.1 \& 7.1.13]{parryang}; \cite[eq.\ 3.14]{baerends1997}; \cite[eq.\ 5.11]{koch2001}; \cite[eq.\ 7.5]{martin2004}; \cite[pg.\ 320]{atkins2011}; \cite[eq.\ 18]{becke2014}; \cite[eq.\ 16.44]{levineQC}.}
\begin{equation}
E[\rho^q,V]=\overline{T}_s[\rho^q]+\sum_{k=1}^{M}\sum_{l>k}^{M} \left(\frac{q_k q_l}{|\vec{r}_k-\vec{r}_l|}\right)+\int{\rho^q(\vec{x}) V(\vec{x}) \  d^3 x}+\frac{1}{2}\int{\frac{\rho^q(\vec{x}) \rho^q(\vec{x}')}{|\vec{x}-\vec{x}'|}\  d^3 x d^3 x'}+E_{xc}[\rho^q]
\ .
\label{energyfunctional3}
\end{equation}
In \eqref{energyfunctional3}, the second and third terms are as in \eqref{energyfunctional2}.  The first term, $\overline{T}_s[\rho^q]$, is the expectation value of the kinetic energy for a new wave function---the minimum energy wave function that has $\rho^q(\vec{x})$ as its charge density taking the Hamiltonian to include only the kinetic energy terms (excluding all interactions):
\begin{equation}
\widehat{H}_s= \sum_{i=1}^{N}\left(\frac{-\hbar^2}{2 m} \nabla_i^2 \right)
\ .
\label{hamiltonianK}
\end{equation}
$\overline{T}_s[\rho^q]$ is usually a good approximation to $\overline{T}[\rho^q]$ and, because the interactions are excluded, it is easier to calculate $\overline{T}_s[\rho^q]$ than $\overline{T}[\rho^q]$.

The fourth term in \eqref{energyfunctional3} is the classical energy associated with electric repulsion within the electron charge density $\rho^q(\vec{x})$, \eqref{electronelectronrepulsionenergy2}, which roughly approximates the expectation value of the energy of electron-electron interaction, $\overline{V}_{ee}[\rho^q]$.  As in section \ref{HFsection}, the presence of this term in \eqref{energyfunctional3} supports Schr\"{o}dinger's idea that electron charge really is spread out with charge density $\rho^q(\vec{x})$.  The third term in \eqref{energyfunctional3} also supports Schr\"{o}dinger's idea, as it is the classical potential energy of Coulomb attraction between the point nuclei and Schr\"{o}dinger's electron charge density, \eqref{electronnucleusenergy2}.

The fifth term in \eqref{energyfunctional3} is a catch-all for the differences between \eqref{energyfunctional2} and \eqref{energyfunctional3},
\begin{equation}
E_{xc}[\rho^q] = \overline{T}[\rho^q] - \overline{T}_s[\rho^q] + \overline{V}_{ee}[\rho^q] - \frac{1}{2}\int{\frac{\rho^q(\vec{x}) \rho^q(\vec{x}')}{|\vec{x}-\vec{x}'|}\  d^3 x d^3 x'}
\ .
\end{equation}
This is called the ``exchange-correlation'' energy functional.  The exact form of this functional is not known, but a variety of impressively successful techniques have been developed to approximate its contribution in \eqref{energyfunctional3}.  Levine \cite[pg. 559]{levineQC} identifies four components of the ground state exchange-correlation energy: ``the \emph{kinetic correlation energy} [which is the difference between $\overline{T}[\rho^q]$ and $\overline{T}_s[\rho^q]$], the \emph{exchange energy} (which arises from the antisymmetry requirement), the \emph{Coulombic correlation energy} (which is associated with interelectronic repulsions), and a \emph{self-interaction correction} (SIC).''

The self-interaction correction is necessary because the fourth term in \eqref{energyfunctional3} treats the electrons together as a single charge density and calculates the energy of electrostatic repulsion within this charge density, erroneously including the energy of electrostatic repulsion within each electron's charge distribution.  In the Hartree-Fock method, this energy of self-repulsion was subtracted out via \eqref{electronselfenergy}.  That equation cannot be used here because it makes explicit use of the electron orbitals and the ground state wave function that has $\rho^q(\vec{x})$ as its associated charge density may not be decomposable into single electron orbitals (put another way, the wave function may not be expressible as a Slater determinant).  Following Kohn and Sham, for any specified charge density one can introduce a Slater determinant wave function that minimizes the energy according to the interaction-free Hamiltonian in \eqref{hamiltonianK}.  It is possible to use the orbitals of this wave function in \eqref{electronselfenergy} to subtract out the self-interaction energy (the Perdew-Zunger self-interaction correction).  However, when approximating $E_{xc}[\rho^q]$ in practice, it is generally better to combine the self-interaction correction with other corrections and to estimate them together.\footnote{For more on the self-interaction correction, see \cite{perdew1981}; \cite[sec.\ 8.3]{parryang}; \cite{baerends1997, baerends2005}; \cite[footnote 33]{argaman2000}; \cite[sec.\ 2.3 \& 6.7]{koch2001}; \cite[sec.\ 4.7]{engel2011}; \cite{tsuneda2014}.}

As in section \ref{HFsection}, density functional theory can be used to calculate the ground state energies and electron structures for the helium atom and the water molecule.\footnote{These applications are discussed in \cite{kim1994}; \cite[sec.\ 8.9]{martin2004}; \cite{baseden2014}; \cite[pg.\ 571]{levineQC}.}  Such calculations would yield charge densities like those pictured in figures \ref{helium} and \ref{water}.  However, they would not yield multi-electron wave functions and would not yield orbitals like those pictured in figure \ref{waterorbitals} (because the ground state wave function is not sought and that not-sought wave function is not assumed to be built from orbitals as a Slater determinant).

In approximating the ground state energy and electron structure of atoms and molecules, we have seen that both the Hartree-Fock method and density functional theory contain clues that electron charge really is spread-out in the way Schr\"{o}dinger suggested.  In both approaches, the energy of electron-nucleus electrostatic interaction is calculated by treating the electrons as a spread-out charge distribution---with the charge density given by \eqref{totalchargedensitypauli}---interacting though classical electrostatic repulsion with the point nuclei.  In both approaches, the energy of electron-electron interaction is calculated by first finding the classical energy of self-repulsion for the electron charge distribution and then including additional quantum correction terms.  Seeing how chemists approximate the underlying quantum physics of atoms and molecules helps to explain why chemists often apply Schr\"{o}dinger's role for amplitude-squared (section \ref{PCsection}) even though Born's role is generally viewed as having superseded it.

\subsection{Further Evidence}\label{FEsection}

Although we have focused in sections \ref{HFsection} and \ref{DFTsection} on the determination of electron structure and ground state energy, there is further support for Schr\"{o}dinger's role in the various ways that his charge density is used in quantum chemistry (where it acts like a true density of charge).  Let us briefly consider a few examples as to how the electron charge density can be put to use once it is calculated (either through density functional theory or through something like the Hartree-Fock method where one first finds the multi-electron wave function and then derives the charge density from it):  The charge density can be used to identify and classify the chemical bonds connecting atoms in molecules \cite{bader1990, shusterman1997, matta2002, bader2013}.  The standard method for calculating the electric dipole moment of a charge distribution in classical electromagnetism is applied to Schr\"{o}dinger's electron charge density when calculating the electric dipole moments of atoms and molecules.\footnote{See \cite[sec.\ 2]{schrodinger1926pt4}; \cite[pg.\ 1066--1068]{schrodinger1926rev}; \cite[sec.\ 3.1]{milonni1976}; \cite[sec.\ 3.4.7]{szaboQC}; \cite{bader1992}; \cite[pg.\ 2]{bader2010}; \cite[sec.\ 4.4]{bacciagaluppi2009}; \cite[sec.\ 14.2]{levineQC}.}  In explaining the equation that he gives for calculating the electric dipole moment of a molecule in his textbook, Levine \cite[pg.\ 407]{levineQC} writes that the equation
\begin{quote}
``is what would be obtained if we pretended that the electrons were smeared out into a continuous charge distribution whose charge density is given by [$\rho^q(\vec{x})$] and we used the classical equation ... to calculate [the electric dipole moment].'' 
\end{quote}
The electron charge density can be used to calculate the effective forces on nuclei via the Hellmann-Feynman electrostatic theorem, which Levine \cite[pg.\ 430]{levineQC} presents as follows:\footnote{See also \cite{feynman1939}; \cite{deb1973}; \cite[pg.\ 40]{milonni1976}; \cite[sec.\ 1.6]{parryang}; \cite[sec.\ 6.2]{gillespie2001}; \cite{bader2003}; \cite[sec.\ 4.1]{bader2010}.}
\begin{quote}
``\emph{the effective force acting on a nucleus in a molecule can be calculated by simple electrostatics as the sum of the Coulombic forces exerted by the other nuclei and by a hypothetical electron cloud whose charge density} [$\rho^q(\vec{x})$] \emph{is found by solving the electronic Schr\"{o}dinger equation.}''
\end{quote}
In addition to forces on nuclei, the electron charge density can be used to calculate forces on atoms within molecules---Ehrenfest forces \cite{bader1990, bader2005, bader2013}.  The electron charge density can also be used to calculate forces between separate atoms and/or molecules.  Levine \cite[pg.\ 460]{levineQC} writes
\begin{quote}
``If the system is a molecule [as opposed to an atom], we view it as a collection of point-charge nuclei and electronic charge smeared out into a continuous distribution.  Electrons are point charges and are not actually smeared out into a continuous charge distribution, but the electronic-charge-cloud picture is a reasonable approximation when considering interactions between two molecules that are not too close to each other.''
\end{quote}
In the three quotes above, Levine describes various appealing features of Schr\"{o}dinger's charge density role for amplitude-squared while refusing to regard it as more than a useful fiction.  In the next section, we will consider the prospects for giving in to temptation.

\section{Charge Density in Quantum Foundations}\label{QFsection}

Interpretations of quantum mechanics aspire to be precise about the ontology (what exists) and the dynamical laws.  Although it has not been a major focus of recent work on quantum foundations, precision about the ontology requires that we be clear about the true distribution of charge.  Do we have point charges, spread-out charge distributions, or a fundamental ontology without electric charge?  In this section, we will discuss the prospects for treating charge as spread out (in the way that Schr\"{o}dinger suggested) within a number of different interpretations of quantum mechanics: Ghirardi-Rimini-Weber (GRW) theory, the many-worlds interpretation, Bohmian mechanics, and two variants of Bohmian mechanics.  The discussion will move quickly as it is primarily aimed at scholars working in the foundations of quantum mechanics.  After discussing these interpretations of non-relativistic quantum mechanics, we will consider extensions of these interpretations to quantum field theory.  The goal in this section is to begin to see how the picture of spread-out charge distributions from the previous two sections might be incorporated into a complete story about the laws and ontology of quantum mechanics (or quantum field theory).  I will not offer final verdicts as to whether such charge distributions should ultimately be included within any particular interpretation.

\subsection{GRW Theory}\label{GRWsection}

GRW theory is a simple example of a spontaneous collapse theory where the ordinary evolution of a quantum wave function under the Schr\"{o}dinger equation is sometimes interrupted by collapse events.  The theory proposes precise laws describing these collapse events where each particle has a constant per-unit-time probability of collapse.  As a consequence of these laws, superpositions that involve large numbers of particles at different locations cannot exist for very long (and thus quantum measurements end up having unique outcomes).  The Schr\"{o}dinger equation and the collapse laws together govern the dynamics of the universal wave function (the wave function for all particles).  In its minimal form (which Tumulka \cite{tumulka2007} and Allori \emph{et al.}\ \cite{allori2008} call GRW$0$), the universal wave function is the full ontology of the theory.  Some have suggested adding to the ontology a ``mass density'' or ``matter density'' (GRWm).\footnote{A third option, GRWf, adds to the ontology special points in spacetime where collapse events (flashes) occur.  One could potentially associate these events with momentarily existing point particles that have both mass and charge.  Or, one could formulate GRWf so that spacetime contains neither mass nor charge.}  Instead of adding a mass density, or in addition, one could consider adding a charge density.  The clues from quantum chemistry presented in section \ref{QCsection} make the inclusion of a charge density appealing.

Taking up an idea from Ghirardi \emph{et al.}\ \cite{ghirardi1995, ghirardi1997},\footnote{See also \cite[pg.\ 38]{goldstein1998}; \cite{maudlin2007}.  Ghirardi \emph{et al.}\ \cite[sec.\ 4.3]{ghirardi1995} consider adding a charge density in addition to a mass density, but reject the idea for reasons that are particular to the continuous spontaneous localization theory that they are discussing (and would not apply to GRW).} Allori \emph{et al.}\ \cite{tumulka2007, allori2008, allori2011, allori2014, goldstein2012} discuss the inclusion of a matter density in GRW theory, calling the resulting variant of GRW ``GRWm.''  They use a simple expression for the matter density, which is the same as Schr\"{o}dinger's expression for the charge density of multiple particles but with particle charges replaced by particle masses (so that for multiple electrons, the expression for the matter density looks just like our earlier expression for the charge density \eqref{totalchargedensitypauli} but with the electron charge $-e$ replaced by the electron mass $m$).  These authors see the inclusion of a matter density as desirable because it allows GRW theory to straightforwardly explain the macroscopic phenomena that we ordinarily take as evidence for quantum mechanics.\footnote{For discussion of this supposed virtue of GRWm, see \cite{maudlin2007, ney2013, neyalbert2013, ney2021}.}  For example, the theory can explain the behavior of a pointer recording the results of a quantum measurement by generating probabilistic predictions as to how the matter composing the pointer will move (in ordinary three-dimensional space).  These authors are not optimistic about attempts to explain such occurrences in quantum theories that only contain the universal wave function (which is often represented as living in high-dimensional configuration space, though the heterodox ``multi-field'' interpretation of the wave function places it in three-dimensional space \cite{forrest1988, belot2012, hubert2018, chen2017, chen2019, romanoF}).  Maudlin \cite[ch.\ 10]{maudlin2011} includes an amusing pair of figures depicting the double-slit experiment in GRWm (with the matter composing the double-slitted barrier, the detection screen, and the particle passing through the slits all visible) and in GRW0 (with nothing at all shown because the theory includes no matter in three-dimensional space, just a quantum wave function).  For the aforementioned project of finding the macroscopic objects of ordinary experience in the ontology of quantum mechanics, there is no obvious need to include a charge density.  However, if we want to understand microphysical reality, the shapes of atoms and molecules, and the relation between quantum mechanics and classical electromagnetism, I think we have good reason to include a charge density in our ontology (see section \ref{QCsection}).

Allori \emph{et al.}\ explain their preference for the title ``matter density'' over ``mass density'' as follows:
\begin{quote}
``Moreover, the matter that we postulate in GRWm and whose density is given by the $m$ function does not \emph{ipso facto} have any such properties as mass or charge; it can only assume various levels of density. For example, the $m$ function is not a source of an electromagnetic field.'' \cite[pg.\ 331--332]{allori2014}
\end{quote}
The authors then go on to consider the idea of endowing this matter with both mass and charge, described by separate mass and charge densities (adopting Schr\"{o}dinger's expression for the charge density), in order to emphasize that this idea is not what they are advocating and ``perhaps not desirable.''  Although the above quote is primarily meant to clarify the ontology that the authors are proposing, it contains an implied criticism of Schr\"{o}dinger's charge density.  There is a concern that if we were to introduce such a charge density, it would not act like a charge density---it would not play the appropriate role in electromagnetic interactions.\footnote{One might also wonder whether the matter density of GRWm acts a source for the gravitational field.  Derakhshani \cite{derakhshani2014} has explored the idea that it does, which would provide support for calling the matter density a ``mass density.''}  In section \ref{QCsection}, we saw that (at least in the context of certain approximations) the electromagnetic interactions between different parts of this charge density can be used to determine and understand the structures of atoms and molecules.  Although we modeled electromagnetic interactions there as unmediated, one could alternatively treat each electron (and proton) as the source of a classical electromagnetic field that is then used to calculate potential energies for the other particles.\footnote{Because there is no electrostatic self-interaction, in section \ref{HFsection} the potential energy of each electron would be determined by only looking at the electromagnetic fields sourced by the other particles.  This raises awkward questions as to whether there are many electromagnetic fields or just one \cite[pg.\ 36, 158]{lange}.}  Given the role that charge density plays in quantum chemistry, the version of GRW that Allori \emph{et al.}\ disavow (including densities of both mass and charge) may be more attractive than their austere GRWm.  Let us call this version ``GRWmc'' (with m for mass and c for charge)\footnote{In addition to densities of mass and charge, GRWmc (and Smc below) should also include densities of momentum and current to describe the flow of mass and charge.} to contrast it with GRWm (with m for mass or matter).\footnote{Another option would be GRWc, with only a charge density and no mass density.  (A potential problem for GRWc is mentioned by Allori \emph{et al.}\ in \cite[footnote 1]{allori2011}.)}

One oddity of GRWmc is that there will be dramatic instantaneous changes in the distribution of mass and charge when the wave function collapses.  Consider, for example, a spin measurement where a single x-spin up electron passes through a z-oriented Stern-Gerlach apparatus and hits a detector screen, at which point its position is entangled with the positions of many other particles (including the particles that compose a pointer indicating where the electron hit), triggering a collapse.\footnote{For reasons that we can put aside here, Stern-Gerlach spin measurements for individual charged particles are difficult and neutral atoms are used instead \cite[pg.\ 230]{ballentine}; \cite[pg.\ 181]{griffithsQM}; \cite[sec.\ 2]{electronsspinmeasurement}.}  According to GRWmc, half of the electron's charge is deflected upwards and half downwards.  Then, when the electron hits the detector its charge jumps to a particular location and the electron's position is recorded.  If you understand this as the electron's charge instantaneously flowing to the position where the electron is found, then we are dealing with infinitely strong momentarily existing currents.  On the other hand, if you understand this as charge disappearing from certain locations and immediately appearing where the electron is found, then you have a violation of local charge conservation (even though, globally, the total amount of charge remains unchanged).  Both options present challenges for extending GRWmc to relativistic quantum physics (though difficulties with relativity are nothing new, and must also be overcome for GRW0 and GRWm \cite{tumulka2007, maudlin2011, bedingham2014}).

\subsection{The Many-Worlds Interpretation}

For the many-worlds interpretation, where the evolution of the wave function is never interrupted by collapse events, it is similarly possible to supplement the wave function with a matter density.  Allori \emph{et al.}\ \cite{allori2008, allori2011} call this Sm (where S denotes that the Schr\"{o}dinger equation governs the evolution of the universal wave function and m again denotes mass or matter density).  As with GRWm, the major motivation for Sm over S0 is that Sm includes matter in ordinary three-dimensional space that can fairly straightforwardly explain the way things appear to us at the macroscopic level.  However, some do not see Sm as superior to S0.  Wallace \cite{wallace2010, wallace2013} thinks that the many-worlds interpretation can perfectly well explain such appearances without adding a matter density (see also \cite{ney2013}; \cite[sec.\ 10.3]{norsen2017}).  Vaidman \cite[sec.\ 3.1]{vaidman2018} thinks that a matter density does not have to be added because it is already there---this density is ``a property of the wave function only'' (see also \cite{lewis2018}).\footnote{One might similarly argue that Schr\"{o}dinger's charge density is already present, deflating the disagreement between S0, Sm, and Smc.}

Sm is just GRWm without the collapse events.  In the Stern-Gerlach spin measurement described above, half of the electron's matter will be deflected upwards and half downwards.  Then, when the electron interacts with the detection screen, the universal wave function will enter a superposition of (i) a piece representing the electron hitting in the upper region and the detector recording ``up'' as the outcome, and (ii) a piece representing the electron hitting in the lower region and the detector recording ``down'' as the outcome.  The matter density generated from this wave function will include a pointer composed of half as much matter as the original in the ``up'' position along with another in the ``down'' position.  If the experimenter had planned to take a step to the left if the result was ``up'' and a step to the right if the result was ``down,'' there will be a half-matter person standing in one location and a half-matter person in the other (each blind to the other's presence).  As Allori \emph{et al.}\ \cite[pg.\ 8]{allori2011} put it, ``Metaphorically speaking, the universe according to Sm resembles the situation of a TV set that is not correctly tuned, so that one always sees a mixture of several channels.''  As more quantum measurements and other similar events occur, the number of channels in the mixture grows and each channel gets dimmer (as each contains less matter).  Still, looking at the matter density associated with each channel you can pick out macroscopic entities like the pointer and the experimenter.  In describing the famous Schr\"{o}dinger cat experiment, Allori \emph{et al.}\ \cite[pg.\ 345--346]{allori2014} explain concisely why they are untroubled by this dimming: it is unobservable.  They write, ``Both cats are there at once, but with reduced mass (which the cats, however, do not notice).''  The cats don't notice their own dimming and they don't notice each other: ``The two cats are, so to speak, reciprocally transparent'' \cite[pg.\ 7]{allori2011}.

Let us now shift from Sm and consider Smc, where we have a universal wave function evolving via the Schr\"{o}dinger equation and also matter possessing densities of mass and charge.  Unlike GRWmc, Smc does not have the problems associated with sudden changes in charge density mentioned at the end of section \ref{GRWsection}.  However, the above-described dimming will result in a progressive loss of mass and charge that might appear problematic (though I think it is acceptable).

To better understand this issue, let us consider a thought experiment along the lines of those that have been raised in the philosophical literature on comparativism about physical quantities \cite{dasgupta2013, dasguptaF, martens2017, martensF2, martensF, baker2021}:  What would happen if at midnight all electric charges became half as strong and all particle masses became half as large?  The answer to this question actually depends on the way that the halving is done.  Suppose that the halving is done by altering the constants that appear in the Schr\"{o}dinger equation, changing the electron charge from $-e$ to $-e/2$ and the electron mass from $m$ to $m/2$ and so on.  If the masses and charges were all to halve in this way, chaos would ensue as stable atoms and molecules would become unstable---though by emitting photons they might relax into a new equilibrium.  Take, for example, the ground state electron wave function for the hydrogen atom,
\begin{equation}
\chi(\vec{x})= \frac{1}{\sqrt{\pi}} \left(\frac{1}{a_0}\right)^{3/2}e^{- |\vec{x}| / a_0} \left(
\begin{matrix}
1 \\
0
\end{matrix}
\right)
\ ,
\label{hyrdogengroundstate}
\end{equation}
where the electron has been chosen (arbitrarily) to be z-spin up and $a_0$ is the Bohr radius, as in \eqref{fexpansion}.  We can assign a size to the hydrogen atom by calculating the root mean square charge radius,
\begin{equation}
\sqrt{\int{ \chi^{\dagger}(\vec{x}) \chi(\vec{x})|\vec{x}|^2 \  d^3 x }} = \sqrt{3} \, a_0
\ ,
\end{equation}
If the masses and charges of the proton and electron that appear in the Schr\"{o}dinger equation are halved, then the ground state electron wave function becomes
\begin{equation}
\chi(\vec{x})= \frac{1}{\sqrt{\pi}} \left(\frac{1}{8 a_0}\right)^{3/2}e^{- |\vec{x}| / 8 a_0} \left(
\begin{matrix}
1 \\
0
\end{matrix}
\right)
\ ,
\end{equation}
which corresponds to a root mean square charge radius of $8\sqrt{3} \, a_0$. The ground state is larger than it was because the electron-nucleus attraction has become weaker (and because the smaller electron mass alters the kinetic energy).  If a hydrogen atom was in its ground state, it will not be when all the masses and charges are halved.  With every atom and molecule in the universe suddenly in an excited state, wild amounts of radiation emission would soon follow.

In Smc, the halving of all masses and charges is commonplace, but it does not happen in the way described above.  Instead, the masses and charges might be halved by performing an appropriate measurement (e.g., measuring the z-spin of an x-spin up particle) and splitting the universal wave function into a superposition of two\footnote{In a realistic scenario, the wave function would branch into many pieces that could be sorted into a z-spin up cluster and a z-spin down cluster \cite[sec.\ 9]{wallace2007}, \cite{wallace2010}, \cite[sec.\ 3.11]{wallaceQM}.  For our purposes here, we can idealize and speak of two branches post-measurement.} equal amplitude branches (worlds).  Given the way that the mass and charge densities are calculated from the wave function, each particle in each branch will only have half of its original mass and half of its original charge.  However, because the reduction in mass and charge comes from a change in the amplitude of the branch and not any alteration of the Schr\"{o}dinger equation, this loss of mass and charge will not affect the behavior of atoms and molecules.  Noting that the reduction in mass and charge has no observable consequences, one could divide the charge density in a branch by the weight of that branch (its amplitude-squared) to arrive at an effective charge density for that branch.\footnote{Carroll \cite[ch.\ 8]{carroll2019} makes similar moves regarding the status of energy in the multiverse as a whole and energy within each branch.}  In Smc, we can thus distinguish between the true charge density (which is conserved across the multiverse but decreases within branches) and the effective charge density for a given branch (which does not decrease as the multiverse grows).

\subsection{Bohmian Mechanics and Other Particle Interpretations}\label{BMsection}

In section \ref{PCsection}, we discussed two potentially viable strategies for reconciling Born and Schr\"{o}dinger's roles for amplitude-squared.  One idea was that the charge of a particle before measurement is spread-out in proportion to psi-squared and that upon measurement (and perhaps at other times) the charge collapses to a fairly well-defined location (as in GRWmc) or at least appears to collapse (as in Smc).  Another idea was that the charge of a particle is always located at a point, but because that point moves rapidly throughout the particle's wave function (spending more time where psi-squared is large), the average charge density over any short interval of time is approximately proportional to psi-squared.

In Bohmian mechanics, the electron is treated as a point particle moving within its wave function according to an explicit, deterministic law of motion: the Bohmian guidance equation.  In general, the electron will not move in the way described above.  For example, in the ground state of the hydrogen atom \eqref{hyrdogengroundstate} the electron will simply stay put at whatever its initial location happens to be.\footnote{See \cite[pg.\ 153]{holland}; \cite[pg.\ 154]{durrtteufel}.}  If we view the electron as a point particle with charge $-e$, then the blurred out (time-averaged) charge density will not even approximately match the amplitude-squared of the electron's wave function (and we will thus be abandoning Schr\"{o}dinger's role entirely, as we are free to do).  Alternatively, one might say that the point electron does not possess charge (a view that has been explicitly defended by Esfeld \emph{et al.}\ in \cite{esfeld2017, esfeld2018, esfeld2020}).  While giving a Bohmian account of the ground state of the hydrogen atom, Holland writes:
\begin{quote}
``Of course, one finds in books informal models of the hydrogen atom in which $|\psi|^2$ is interpreted as a `charge density' in a `cloud of probability', with the implication that the electron is somehow `everywhere' that the probability is finite. Indeed, since for us the wave is a component of the electron, it is true that the latter is extended, but we also have a corpuscle [particle] that occupies a definite position in space.  The `charge density' is then both a characteristic of the physical field [the wave function] and an indication of the likely position of the corpuscle.'' \cite[pg.\ 155]{holland}
\end{quote}
After rejecting the kind of model of the electron that we find in GRWmc and Smc, Holland suggests that `charge density' is a property of the wave function.  Depending on how seriously we take the scare quotes, this might mean either that the physical charge is really spread-out in proportion to amplitude-squared (in which case one begins to question why the particle is there) or, more plausibly, that at the fundamental level there is no electric charge ($-e$ is just a constant in the Schr\"{o}dinger equation).

The dynamics of Bohmian mechanics can be modified so that particle trajectories deviate randomly and significantly from the standard Bohmian trajectories.  The existent proposals along these lines are not seeking a match (and do not achieve a match) between the time-averaged charge density that results from particle motion and the amplitude-squared of the wave function.\footnote{See \cite{nelson1985, goldstein1987, bohmhiley1989}; \cite[pg.\ 645]{nelson1990}; \cite{bacciagaluppi1999}; \cite[footnote 5]{gao2018}.}  However, by forgoing continuous trajectories and having particles jump randomly from one location to another, Gao \cite{gao2014, gao2017, gao2018, gao2020} has sought to reconcile Born and Schr\"{o}dinger's roles in this way.\footnote{Gao \cite[ch.\ 8]{gao2017} combines this ``random discontinuous motion of particles'' with a collapse dynamics for the wave function.  Without this wave function collapse, his proposal would be a many-worlds theory: Bell's Everett (?) theory \cite{bell1981}; \cite[sec.\ 5.1]{barrett1999}; \cite[sec.\ 6.2]{allori2008}; \cite[sec.\ 4]{allori2011}; \cite{maudlin2016}; \cite[ch.\ 8]{gao2017}.}  The idea is that the configuration of particles changes randomly so that the integral of the amplitude-squared of the wave function over a region of configuration space at some time $t$ gives the fraction of time that the actual configuration of particles spends in that region within any short time interval centered at $t$.  Gao points out an important virtue of this strategy for reconciling the two roles: if Schr\"{o}dinger's charge density is only a time-averaged charge density for point charges, that would neatly explain the absence of electrostatic self-interaction between different parts of a single electron's cloud (something that was unexplained in sections \ref{HFsection} and \ref{DFTsection}).  Gao's approach also gives a nice explanation of exchange and correlation energies in atoms and molecules: because of both Pauli exclusion and electrostatic repulsion, electrons avoid spending time near one another.  Thus, on Gao's approach, the (time-averaged) potential energies of atoms and molecules---as calculated from the Hamiltonian in \eqref{hamiltonian}---can be understood entirely in terms of classical electrostatic attraction and repulsion (the contributions that were called ``quantum corrections'' in sections \ref{HFsection} and \ref{DFTsection} would have classical explanations). 

There is another variant of Bohmian mechanics that is worth mentioning.  This interpretation has been called ``Newtonian quantum mechanics'' \cite{sebens2015} or ``many interacting worlds'' \cite{HDW}.\footnote{Similar approaches are discussed in \cite{holland2005, tipler2006, schiff2012, bostrom2015, bokulich2020}.}  According to this interpretation, there are a vast number of worlds and in each world particles have precise locations.  At the fundamental level, there is no wave function (though a wave function can be introduced as a non-fundamental description of the worlds, with its amplitude-squared giving the density of worlds in configuration space).  Particles feel ordinary classical forces (such as electrostatic attraction and repulsion) from other particles in the same world and quantum forces from interactions between worlds.  Focusing on the simple case of a lone helium atom, there could be a multitude of worlds that all have the atom's nucleus located near the origin and each have the two electrons at different locations---so that if you plotted the electron density across worlds you would end up with the familiar electron cloud shown in figure \ref{helium}.  If each electron has charge $-e$, then this electron density is a charge density.  The total electron charge (across all worlds) would be absurdly large and the electron charge would not be spread-out within any world, but still a charge density like Schr\"{o}dinger's would be present according to this interpretation.\footnote{It would be unwise to assign each electron a smaller charge so that the total electron charge (across all worlds) for the helium atom is what you'd expect, $-2e$.  One problem is that the branching that occurs upon measurement within this interpretation is a process where clusters of worlds separate from one another (in configuration space), leading to a dimming like that discussed for the many-worlds interpretation.  For example, if you decide to move your helium atom (with total electron charge $-2e$) left or right depending on the result of a quantum measurement with two equiprobable outcomes, then once that is done the total electron charge of the atom will be $-e$ (not $-2e$) in the cluster of worlds where the atom was moved to the left and will also be $-e$ (not $-2e$) in the cluster where the atom was moved to the right.}  The ground state potential energy of an atom or molecule can be viewed here as an average of electrostatic attractions and repulsions across different real worlds where each electron, proton, and neutron has a definite location.  This would neatly explain the absence of electrostatic self-interaction (as electrons are not spread-out within any world) and also the presence of exchange and correlation energies (as the density of worlds in configuration space drops for configurations where electrons are close to one another, because of both Pauli exclusion and electrostatic repulsion).

\subsection{Quantum Field Theory}\label{QFT}

Thus far, we have focused on interpretations of non-relativistic quantum mechanics.  Going deeper, we would like to find interpretations of relativistic quantum field theory that are clear about the laws and ontology.  To get started on such an endeavor, we must ask what replaces the wave function over particle configurations as a (perhaps only partial) representation of the physical state.  One option is to stick with a wave function over configuration space and then either expand the configuration space so that the wave function assigns amplitudes to different numbers of particles being in different configurations, or, alternatively, keep the true number of particles fixed and argue that particle creation and annihilation is only apparent (viewing the creation of an electron-positron pair as an electron exiting the Dirac sea).\footnote{See \cite{struyve2011, tumulka2018, durr2020} and references therein.}  If we retain wave functions over particle configurations in quantum field theory, then the ideas in sections \ref{GRWsection}--\ref{BMsection} as to how one might vindicate Schr\"{o}dinger's charge density role for amplitude-squared could potentially be extended to quantum field theory without radical modification.  Another option for replacing the wave function over particle configurations from non-relativistic quantum mechanics is to introduce a wave functional over field configurations, treating quantum field theory as a theory describing superpositions of classical field states in the same way that non-relativistic quantum mechanics describes superpositions of classical particle states.  This view of quantum field theory as a true theory of fields alters the landscape of options for preserving Schr\"{o}dinger's idea of a spread-out charge density.  Let us take a moment to see how.

For electrons, the classical field that would be viewed as entering quantum superpositions is the Dirac field: a four-component complex-valued\footnote{Ultimately, it may be best to view classical fermion fields (like the Dirac field) as Grassmann-valued, not complex-valued.  The reasons for this and the challenges that accompany such a picture will not be discussed here (see \cite{floreanini1988, jackiw1990, hatfield, valentini1992, valentini1996, struyve2010, struyve2011, positrons}).  The problems with Grassmann field values have led some to advocate only using wave functionals for bosons \cite{bohm1987, bohmhiley}.} field $\psi$ that carries both energy and charge.  For atoms or molecules, the classical Dirac field could be in various different states with clouds of charge density like those in figures \ref{helium} and \ref{water}.  Thus, at the classical level, electrons are being described as spread-out charge densities (roughly in the way that Schr\"{o}dinger pictured), not as point charges.\footnote{I have recently studied this classical field description of electrons in a series of papers focused on understanding electron spin as the actual rotation of energy and charge \cite{howelectronsspin, smallelectronstates, positrons, electronsspinmeasurement}.}  In quantum field theory, we then have a superposition of such classical field states from which we can determine the actual electron charge density in a number of ways depending on our preferred interpretation of quantum field theory: (i) we could use the amplitude-squared of the wave functional to take a weighted average of the charge densities for different possible classical states of the Dirac field (in extensions of GRWmc or Smc), (ii) we could add to the ontology a single actual Dirac field configuration with a definite charge density (in an extension of Bohmian mechanics\footnote{See \cite[ch.\ 4]{valentini1992}; \cite{valentini1996, struyve2010, struyve2011}.}), (iii) we could add an actual Dirac field configuration that changes rapidly and use the average of the field's charge density over a short period of time as the relevant charge density (in an extension of Gao's approach from section \ref{BMsection}), (iv) we could replace the wave functional with a vast number of worlds, where in each world the Dirac field has a definite configuration and thus a definite charge density, which could be analyzed on its own or averaged across some or all of the worlds (in an extension of the many interacting worlds approach from section \ref{BMsection}).  It would require further research to see whether any or all of these proposals accurately reproduce Schr\"{o}dinger's charge density (calculated using non-relativistic quantum mechanics via the methods of section \ref{QCsection}) when you approximate quantum field theoretic descriptions of the states of atoms and molecules.  But, it is clear that on any of these proposals, the negative electron charge of an atom or molecule would be viewed as spread-out and that the fundamental ontology would not include point charges.

\section{Conclusion}

Schr\"{o}dinger proposed that the amplitude-squared of the quantum wave function gives the density of charge, as described in section \ref{PCsection}.  Within quantum chemistry, this idea remains popular despite its tension with Born's idea that the amplitude-squared of the wave function tells us about the probability for finding particles in different locations.  In section \ref{QCsection}, we examined some of the reasons why it is appealing to think of electron charge as spread out in the way Schr\"{o}dinger proposed.  In both the Hartree-Fock method and density functional theory, Schr\"{o}dinger's charge density allows us to interpret two important contributions to the ground state potential energies of atoms and molecules as classical contributions from Coulomb attraction (between the nuclei and the electron charge distribution) and Coulomb repulsion (within the electron charge distribution).  Although quantum chemists regularly treat wave functions as describing spread-out distributions of charge, scholars working on the foundations of quantum mechanics rarely explicitly include such charge densities in the ontologically precise formulations of quantum mechanics that they propose.  But, charge densities can be incorporated into a number prominent approaches to the foundations of quantum mechanics.  In section \ref{QFsection}, we saw how.  More research is needed to determine whether the variants of these approaches that include spread-out distributions of charge are superior to the alternatives.  Here I have argued that their fit with quantum chemistry is a point in their favor.  When we move to quantum field theory, I think the case for a spread-out electron charge density is particularly strong as the theory can be viewed as describing quantum superpositions of classical field states where electron charge is spread out.

\vspace*{12 pt}
\noindent
\textbf{Acknowledgments}
Thank you to Craig Callender, Eddy Keming Chen, Scott Cushing, Maaneli Derakhshani, Mario Hubert, Joshua Hunt, Gerald Knizia, Logan McCarty, and Eric Winsberg for helpful feedback and discussion.

\end{document}